\documentclass[lettersize,journal]{IEEEtran}
\usepackage{amsmath,amsfonts}
\usepackage{algorithmic}
\usepackage{algorithm}
\usepackage{array}
\usepackage[caption=false,font=normalsize,labelfont=sf,textfont=sf]{subfig}
\usepackage{textcomp}
\usepackage{stfloats}
\usepackage{url}
\usepackage{verbatim}
\usepackage{graphicx}
\usepackage{cite}
\usepackage{hyperref}
\usepackage{multirow}
\usepackage{graphicx} 
\usepackage{booktabs} 
\usepackage{makecell}
\usepackage{tabularx}
\usepackage{amssymb}
\usepackage{enumitem}
\usepackage{svg}
\svgsetup{inkscapelatex=false}

\usepackage{xcolor}
\usepackage{listings}
\usepackage[utf8]{inputenc}

\usepackage{tcolorbox}
\tcbuselibrary{breakable, skins}

\usepackage{mathptmx}
\usepackage[normalem]{ulem}  

\usepackage[switch]{lineno}  
\newcommand{\revise}[1]{#1} 
\newcommand{\mrevise}[1]{#1} 

\newtheorem{definition}{Definition}
\newcolumntype{S}{>{\centering\arraybackslash\hsize=0.7\hsize}X}
\newcolumntype{T}{>{\centering\arraybackslash\hsize=0.1\hsize}X}  

\newcolumntype{M}{>{\raggedright\arraybackslash\hsize=0.9\hsize}X}

\newcolumntype{L}{>{\raggedright\arraybackslash\hsize=1.4\hsize}X}

\newcolumntype{Y}{>{\centering\arraybackslash}X}

\newcolumntype{A}{>{\raggedright\arraybackslash\hsize=0.8\hsize}X}
\newcolumntype{B}{>{\centering\arraybackslash\hsize=0.2\hsize}X}

\newtcolorbox{promptbox}[1]{ 
	colback=gray!5!white,     
	colframe=gray!75!black,   
	fonttitle=\bfseries,      
	title=#1,                 
	arc=2mm,                  
	outer arc=2mm,            
	left=2mm,                 
	right=2mm,                
	top=1mm,                  
	bottom=1mm,               
	boxrule=0.8pt,            
	breakable=true,           
	fontupper=\small, 		  
}

\begin{document}
	

\title{Enhancing Cloud Network Resilience via a Robust LLM-Empowered Multi-Agent Reinforcement Learning Framework}

\author{Yixiao~Peng,~
	Hao~Hu,~
	Feiyang~Li,~
	Xinye~Cao,~
	Yingchang~Jiang,
	Jipeng~Tang,~
	Guoshun~Nan,~\IEEEmembership{Member,~IEEE},~
	and~Yuling~Liu,~\IEEEmembership{Member,~IEEE}
\thanks{This work was supported in part by the National Natural Science Foundation of China under Grants 61902427 and 62471064, and by the National Key Research and Development Program of China under Grants 2023YFC3306305 and 2022YFB2902200. (Corresponding author: Yuling Liu.)}
\thanks{Y. Peng, H. Hu, F. Li, Y. Jiang, and J. Tang are with the State Key Laboratory of Mathematical Engineering and Advanced Computing, Zhengzhou, China, and also with the Henan Key Laboratory of Information Security, Zhengzhou, China (e-mail: 13730818683@163.com; wjjhh\_908@163.com; lfyxxgcdx@163.com; yingchangjiang@163.com; 19138037519@163.com).}
\thanks{X. Cao and G. Nan are with the National Engineering Research Center for Mobile Network Technologies, Beijing University of Posts and Telecommunications, China (e-mail: caoxinye@bupt.edu.cn; nanguo2021@bupt.edu.cn).}
\thanks{Y. Liu is with the Institute of Information Engineering, Chinese Academy of Sciences, Beijing, China (e-mail: liuyuling@iie.ac.cn).}%
\thanks{Yixiao Peng and Hao Hu are equally contributed.}
\thanks{This work has been submitted to the IEEE for possible publication. Copyright may be transferred without notice, after which this version may no longer be accessible.}
\vspace{-2.5em}
}

\markboth{Journal of \LaTeX\ Class Files,~Vol.~14, No.~8, August~2021}%
{Shell \MakeLowercase{\textit{et al.}}: A Sample Article Using IEEEtran.cls for IEEE Journals}


\maketitle

\begin{abstract}
	While virtualization and resource pooling empower cloud networks with structural flexibility and elastic scalability, they inevitably expand the attack surface and challenge cyber resilience. Reinforcement Learning (RL)-based defense strategies have been developed to optimize resource deployment and isolation policies under adversarial conditions, aiming to enhance system resilience by maintaining and restoring network availability. However, existing approaches lack robustness as they require retraining to adapt to dynamic changes in network structure, node scale, attack strategies, and attack intensity. Furthermore, the lack of Human-in-the-Loop (HITL) support limits interpretability and flexibility. To address these limitations, we propose CyberOps-Bots, a hierarchical multi-agent reinforcement learning framework empowered by Large Language Models (LLMs). Inspired by MITRE ATT\&CK’s ``Tactics-Techniques'' model, CyberOps-Bots features a two-layer architecture: (1) An upper-level LLM agent with four modules—ReAct planning, IPDRR-based perception, long-short term memory, and action/tool integration—performs global awareness, human intent recognition, and tactical planning; (2) Lower-level RL agents, developed via heterogeneous separated pre-training, execute atomic defense actions within localized network regions. This synergy preserves LLM adaptability and interpretability while ensuring reliable RL execution. Experiments on real cloud datasets show that, compared to state-of-the-art algorithms, CyberOps-Bots maintains network availability 68.5\% higher and achieves a 34.7\% jumpstart performance gain when shifting the scenarios without retraining. \mrevise{To our knowledge, this is the first study to establish a robust LLM-RL framework with HITL support for cloud defense.}
\end{abstract}


\section{Introduction}
\label{sec:introduction}

\begin{figure}[htbp]
	\centering
	\includegraphics[width=\columnwidth]{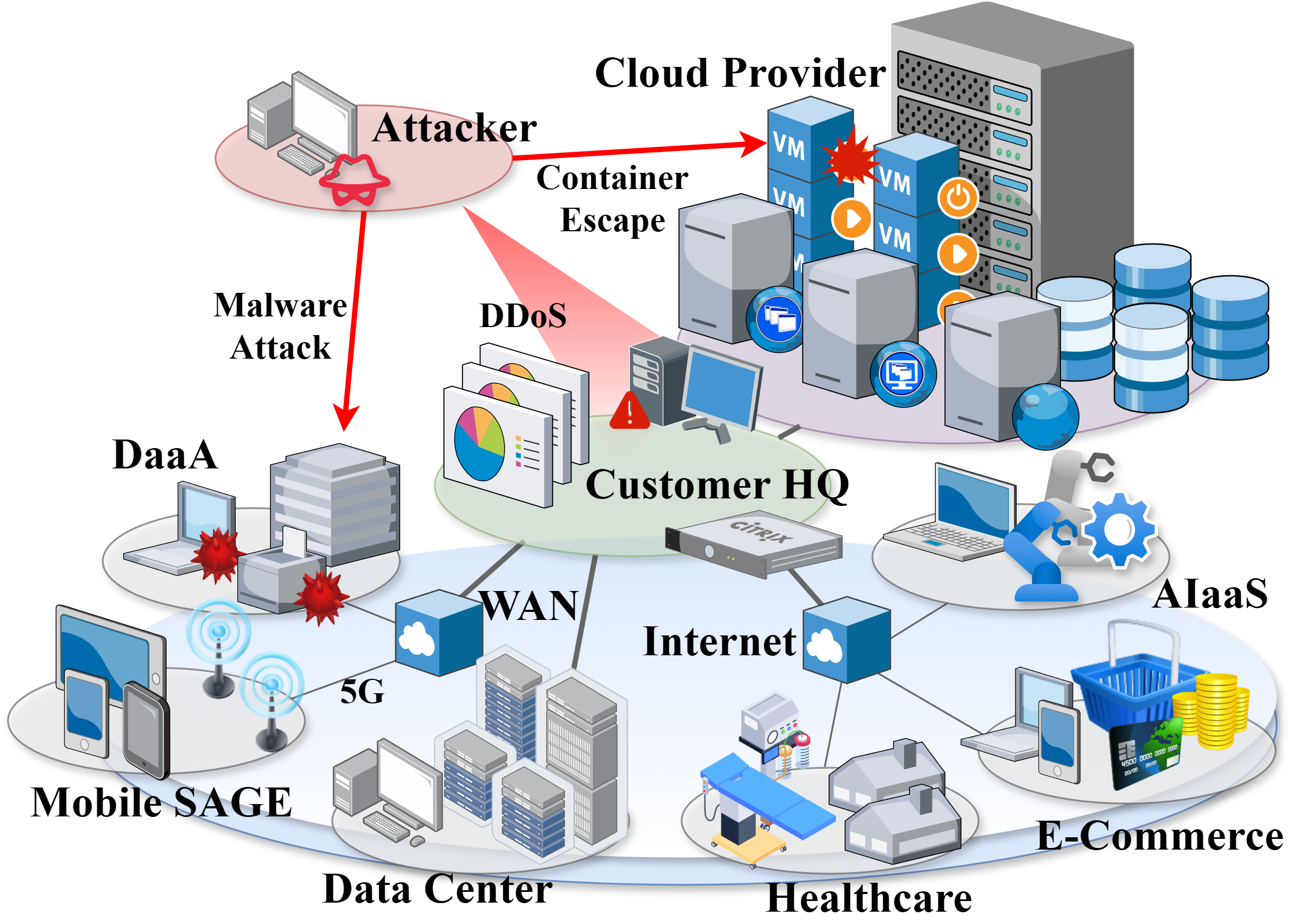} 
	\caption{While technologies like virtualization and elastic scaling provide dynamic flexibility, they inevitably expand the attack surface. This increased exposure facilitates specific cloud attacks, such as container escape and malware in cloud storage, thereby challenging system resilience.}
	\label{fig:intro}
\end{figure}

With the widespread adoption of cloud computing technologies, cloud networks have become critical infrastructure supporting various information services\cite{CloudAdvancements, ebpf}. \revise{Their applications have penetrated numerous sectors including government, finance, industry, and healthcare\cite{CloudAcceleration, autonomousCloud}. Cloud networks exhibit highly dynamic characteristics: virtualization and elastic scaling mechanisms lead to frequent creation, destruction, and reconfiguration of network components like virtual machines and container instances, resulting in continuous changes in network structure and scale\cite{CloudAcceleration}. While enhancing resource utilization efficiency and system flexibility, cloud networks also present a large-scale, distributed attack surface, making them ideal targets for coordinated attacks and posing severe challenges to network resilience\cite{dynAttack}.} Consequently, Reinforcement Learning (RL) and Deep Reinforcement Learning (DRL) have been widely applied in cyber security decision-making due to their ability to autonomously learn optimal defense strategies through environmental interaction\cite{NguyenDeep, FoleyAutonomous}. The core principle involves modeling the attack-defense process as a Markov Decision Process and solving for the optimal policy by maximizing cumulative rewards, thereby balancing defense costs with network availability to enhance cyber resilience. However, the dynamic nature of cloud networks and the threats they face impose adaptability challenges on existing DRL-based network defense decision-making methods.


\begin{figure*}[htbp]
    \centering
    \includegraphics[width=\textwidth]{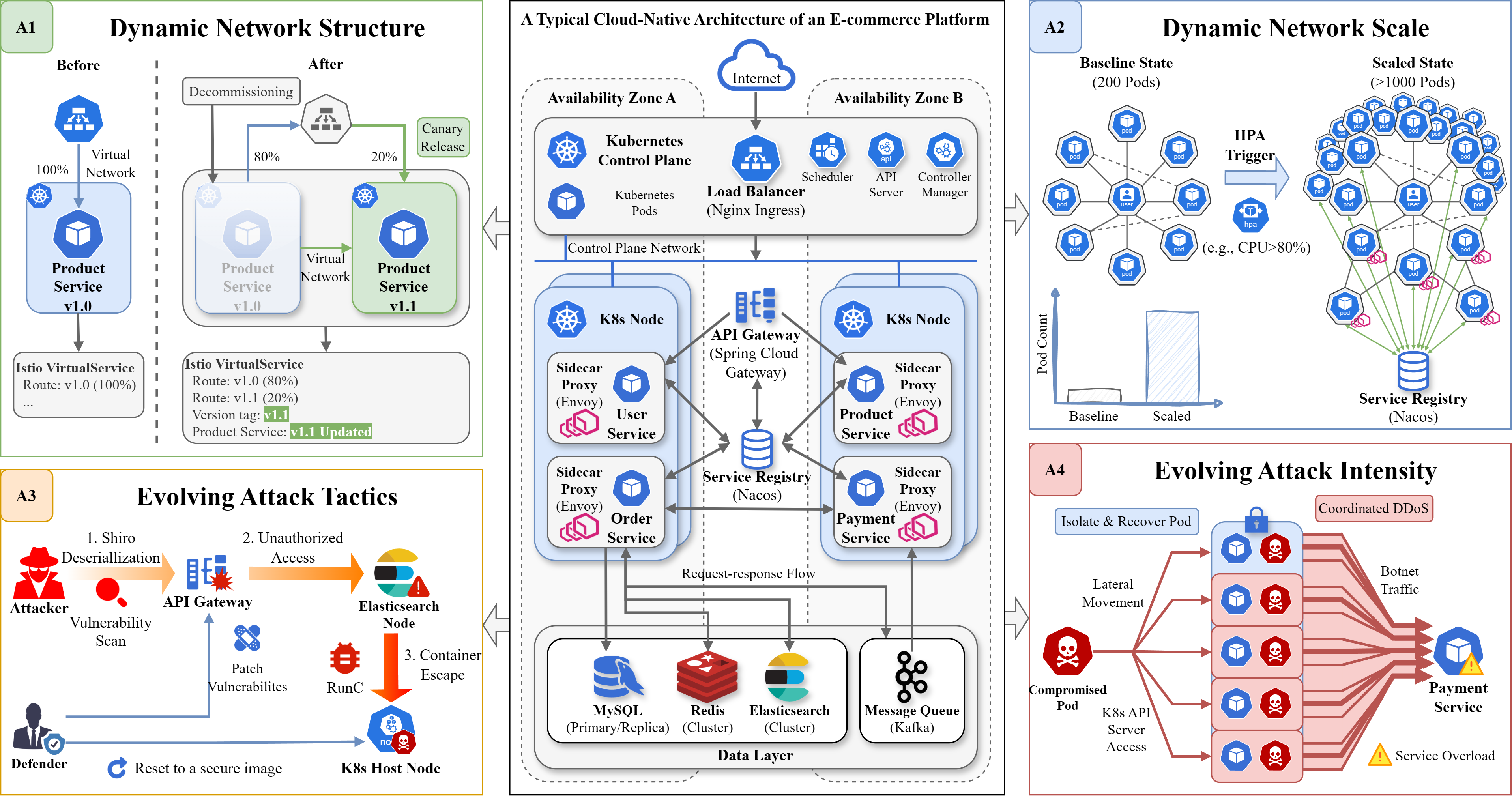}
    \caption{A typical cloud-native e-commerce architecture, exemplifying the four dynamic aspects (A1-A4). i) The network frequently performs elastic scaling and instance migration in response to workload fluctuations. ii) Business instances dynamically expand as new features are deployed or partners integrated. iii) Meanwhile, attack tactics and scale are constantly evolving, from initial port scanning of public entry points, to compromising databases via vulnerable middleware, and ultimately escalating into coordinated DDoS attacks that cripple core services.}
    \label{fig:cloud}
\end{figure*}

Specifically, as illustrated in Fig.\ref{fig:cloud}, the cloud-native architecture of a large e-commerce platform serves as a typical example of a cloud network. \mrevise{It consists of virtual machine clusters across multiple Availability Zones, containerized microservices, load balancers, and distributed database instances. Due to fluctuating business loads, this network frequently undergoes elastic scaling and instance migration, leading to continuous changes in the virtual network topology and east-west traffic policies (A1).} Concurrently, business modules dynamically scale the number of instances as new features are deployed or partners are integrated, resulting in a highly dynamic number of network nodes (A2). Meanwhile, attack tactics and scales are constantly evolving: an attack may start with port scanning and service probing on public-facing entry points, then compromise database instances through vulnerable middleware (A3)\cite{HuOptimal, vyas2023automated}, and finally escalate into coordinated attacks aiming to paralyze core services via encryption ransomware or DDoS flooding (A4)\cite{LiDynPen}. To counter such threats, defenders can take various actions, such as resetting compromised instances to clean images, isolating compromised container nodes to contain spread, patching microservice vulnerabilities to eliminate attack surfaces, or restoring isolated nodes to maintain business continuity. 

\revise{Therefore, we define a \textbf{robust cloud network defense decision-making framework} as one that can seamlessly adapt to four key dynamic aspects (A1-A4) while maintaining defense effectiveness without the need for retraining:}

\begin{itemize}
    \item \textbf{A1: Adaptability to Network Structure.} Alibaba Cloud cluster tracking data shows that 11,089 container rescheduling and co-locating events can occur within 12 hours\cite{AliTrace}, indicating extremely high network structure dynamics. Therefore, when network configurations and node connections change, the defense decision-making framework should seamlessly adapt without retraining.
    
    \item \textbf{A2: Adaptability to Network Scale.} According to Google Cloud cluster statistics, the node scale can expand to over ten thousand\cite{reiss2012heterogeneity}. Therefore, the defense mechanism should effectively scale to networks containing a large number of diverse node types, avoiding state space explosion.
    
    \item \textbf{A3: Adaptability to Attack Policy.} The Capital One data breach incident\cite{novaes2020case} combined multiple techniques such as Server-Side Request Forgery (SSRF) vulnerabilities, AWS instance metadata service abuse, IAM credential theft, and S3 bucket enumeration, reflecting the multi-stage nature of cloud attacks. Therefore, the defense system should dynamically respond to diverse, multi-stage attack strategies.
    
    \item \textbf{\mrevise{A4: Adaptability to Attack Scale.}} Google Cloud Armor statistics indicate that attack traffic targeting a single user can originate from 5,256 source IPs across 132 countries, demonstrating extremely high concurrency\cite{google_cloud_armor_how_2022}. Therefore, when the number of concurrent attackers increases, the defense strategy should maintain robustness and simultaneously handle multiple attack paths.
\end{itemize}

\revise{Here, we clarify the specific meaning of ``no retraining'' in the context of this work. Specifically, this term indicates that the framework does not require online gradient updates or parameter fine-tuning during inference to maintain defensive performance when encountering the aforementioned dynamics.}

Existing research struggles to simultaneously meet the aforementioned adaptability requirements due to two root causes. First, current defense models typically encode network states into fixed-dimensional vectors. For instance, Purves et al.\cite{purves2024causally} integrate network attributes such as node connectivity, vulnerability CVSS scores, and critical assets into a fixed-length vector, causing the internal parameters (e.g., neural network input tensor shape, weights) to be deeply coupled with the specific network scale (A2) and structure (A1) during training. Once the network environment changes, the state vector dimensionality alters, often necessitating resource-intensive retraining. Second, existing approaches usually employ similar or identical attack patterns and scales during both training and testing phases. \mrevise{For example, \cite{purves2024causally, ZhuLearning, XiaoRLAPT} utilize exactly the same attack policies in both phases. Consequently, when confronted with unseen, phase-evolving attack policies (A3) or concurrent attacks of significantly larger scale (A4), the models fail to generalize effectively.}

Given this, we propose CyberOps-Bots (Cyber Operation Bots), a robust hierarchical multi-agent framework based on Large Language Models (LLMs) and RL, designed to build an agent cluster for automated collaborative cyber defense operations in cloud networks. Inspired by the MITRE ATT\&CK framework \cite{ATTCK}, the framework adopts a ``Tactics-Techniques'' co-governance design: the upper-layer LLM agent is responsible for global situation awareness, human intent recognition, defensive tactical planning, and resource scheduling, leveraging its semantic understanding and multi-step reasoning capabilities; the lower-layer RL agents execute specific defensive actions within localized network regions, relying on their RL mechanisms for efficient and precise control. Through this hierarchical design, the framework preserves the adaptability and interpretability of LLMs in high-level planning while ensuring the reliability and executability of low-level actions via RL. \revise{Furthermore, the framework inherently supports human-in-the-loop (HITL), allowing security experts to inject prior knowledge or intervene in real-time through natural language at the LLM tactical layer, enabling human-machine collaborative defense.} The decision-making process is recorded as an auditable, readable tactical planning reasoning chain. By seamlessly integrating human expert domain knowledge and judgment into the automated decision cycle, the system can flexibly adjust strategies in complex, dynamic adversarial environments, compensating for the limitations of purely algorithmic models in semantic understanding, ethical trade-offs, and response to unforeseen events.

\revise{It is acknowledged that a centralized LLM planner, while offering interpretability and tactic planning, may introduce new security considerations in deployment, such as becoming a single point of failure or a target for adversarial prompt manipulation and observation poisoning. Addressing these model-level security threats, however, extends beyond the scope of this paper, which focuses on establishing the architectural paradigm and validating its core adaptive capabilities.}

The main contributions of this paper are as follows:

\begin{itemize}
    \item We propose a natural language-based method for representing cyber adversarial scenarios. By converting high-dimensional, structured network state information and network context into natural language descriptions as input for LLM-based global situational awareness, the defense framework becomes independent of specific network structures. This enables seamless adaptation to changes in network topology (A1) and scale (A2).
    \item We design a hierarchical decision-making framework combining LLM and RL suitable for large-scale network environments. The upper-layer LLM agent performs global tactical planning using the ReAct paradigm, while lower-layer RL agents execute atomic defense actions within localized network regions. This effectively mitigates the state space explosion problem caused by network scaling (A2).
    \item We introduce a heterogeneous separated pre-training architecture. By designing specialized reward functions and training scenarios, a set of functionally heterogeneous RL expert agents are trained offline. The LLM then performs semantic-level tactical parsing and planning based on real-time attack-defense situations, scheduling different RL agents via tool invocation to form adaptive defense strategies tailored to \mrevise{varying attack policies (A3)}.
    \item We develop a long-short term memory mechanism for multi-stage, multi-path attacks. This mechanism stores multiple attack chains and retrieves them based on similarity matching, allowing the LLM to track the evolution of multiple attack chains across time steps and conduct defense using “reactive” memory and tool calls. This enhances the framework’s capability to handle increased concurrent attack scales (A4).
\end{itemize}

\section{Related Works}

\begin{table*}[htbp]
	\footnotesize 
	\caption{Comparison of Existing Defense Methods Against Robustness Requirements.}
	\label{tab:works}
	\centering
	
	
	\begin{tabularx}{\textwidth}{l cccc S S S L L M}
		\hline
		\multirow{2}{*}{Ref.} & \multicolumn{4}{c}{Robust Adaptability} & \multirow{2}{*}{Framework} & \multirow{2}{*}{Scenario} & Resilience & \multirow{2}{*}{Attack Policy} & \multirow{2}{*}{Interpretability} & \multirow{2}{*}{HITL Support} \\ \cline{2-5}
		& A1 & A2 & A3 & A4 &  &  & Aware &  &  & \\ \hline
		
		\cite{peng2025llm4game} & $\bullet$ & $\circ$ & $\bullet$ & $\bullet$ & LLM-RL-KI & Cloud storage & Yes & Multi-stage Rule-Based Policies & Based on knowledge vector (Explicit) & Auditable Knowledge Quadruple \\
		
		\cite{li2025idsagent} & $\bullet$ & $\bullet$ & $\bullet$ & $\circ$ & LLM Agent & IoT & No & - & Based on LLM ReAct reasoning (Explicit) & Customizable Workflow \\
		
		\cite{LiDual} & $\bullet$ & $\circ$ & $\bullet$ & $\circ$ & OAPP & 5G-ICPS & No & Dual RL Agents & Based on strategy output (Implicit) & No \\
		
		\cite{XiaoRLAPT} & $\bullet$ & $\bullet$ & $\circ$ & $\circ$ & Actor-Critic RL & Smart grids & Yes & Trained Q-Learning & Based on strategy output (Implicit) & No \\
		
		\cite{singh2024hierarchical} & $\circ$ & $\bullet$ & $\bullet$ & $\circ$ & Hierarchical PPO & Cloud computing & Yes & Multiple Static Rule-Based Policies & Interpretable metrics (Implicit) & No \\
		
		\cite{GMANRLD} & $\circ$ & $\circ$ & $\bullet$ & $\bullet$ & GMADRLD & Cloud computing & No & Multiple Policies with Variants & Based on strategy output (Implicit) & No \\
		
		\cite{purves2024causally} & $\bullet$ & $\circ$ & $\bullet$ & $\circ$ & PPO+SCM & Autonomous cyber operation & Yes & Multiple Static Rule-Based Policies & Based on SCM rewards (Implicit) & Limited: SCM Human Knowledge Integration \\
		
		\cite{CaoDeepMTD} & $\circ$ & $\bullet$ & $\bullet$ & $\circ$ & PPO & SDN & Yes & Multiple Static Rule-Based Policies & Based on strategy output (Implicit) & No \\
		
		\cite{RenMultiagent} & $\circ$ & $\circ$ & $\bullet$ & $\circ$ & DDQN & IoT & No & DQN Agent & Based on strategy output (Implicit) & No \\
		
		Ours & $\bullet$ & $\bullet$ & $\bullet$ & $\bullet$ & LLM+RL Agents & Cloud network & Yes & Multi-stage Intensity-Escalation Attack & Based on LLM ReAct reasoning (Explicit) & Expert Instructions Integration \\ \hline
	\end{tabularx}
\end{table*}

\revise{Cloud networks, characterized by dynamic scheduling and scalability, face significant threats from multi-staged attacks. While technical approaches exist, they struggle to simultaneously adapt to changes in network structure (A1), scale (A2), attack policy (A3), and intensity (A4).}

\revise{RL-based methods model cyber-defense as MDPs or game-theoretic processes \cite{NguyenDeep, FoleyAutonomous, JainRecent, rass2023game, KongOptimal} and learn strategies within the models through RL algorithms. These have been applied to CPS \cite{MohamedResilient, bitirgen2024markov}, edge computing \cite{ramana2022ambient}, and IoT \cite{SoussiMoving, HouEnhancing}. To enhance decision-making, researchers have employed MARL \cite{tang2024method, McdonaldFinding} and MADRL \cite{ZhuLearning, ZhangGame} against APTs. However, these methods often require retraining or suffer from policy failure when network scales expand or attack phases shift.}

To address the adaptability limitations, numerous works have proposed adaptive defense frameworks.

\revise{In terms of enhancing adaptability to attack strategies, Ren et al.\cite{RenMultiagent} employed adversarial training and synchronous interaction mechanisms. While promising, their experimental validation focused on fixed network and attacker scales. Cao et al.\cite{CaoDeepMTD} introduced a self-evolving Moving Target Defense (MTD) approach aimed at overcoming the generalization challenges arising from discrepancies between training environments and attack-defense scenarios. However, evaluations of this method were mainly conducted on a fixed Software-Defined Networking (SDN) topology. Purves et al.\cite{purves2024causally} incorporated a causal inference model to replace traditional reward models. Nevertheless, the computational complexity of reward based on causal graphs may pose scalability challenges in large-scale networks. Targeting multi-stage APT attacks in cloud computing networks, Chen et al.\cite{GMANRLD} proposed the GMADRLD algorithm, which enhances resource allocation efficiency by grouping agents. Experiments demonstrated its adaptability to varying attack strategies and scales.}

In terms of enhancing adaptability to network scale and structure, Singh et al.\cite{singh2024hierarchical} employed a hierarchical PPO with a two-stage training method to address the state space explosion problem in decentralized networks. It is worth noting that the training of upper-layer agents in this framework incorporates expert rules, which may involve specific manual adjustments when adapting to different cyber-defense scenarios. \revise{While traditional hierarchical RL methods\cite{HRL} decompose complex tasks into sub-goals, their high-level policies are often constrained by fixed numerical state-action mappings and struggle with cross-topology generalization. In contrast, our LLM-based planner shifts the hierarchical coordination from the numerical level to the semantic level.} \mrevise{Xiao et al.\cite{XiaoRLAPT} focused on APT defense in large-scale smart grids. While their approach demonstrates capability in large-scale node management (each MDMS node connects 50–150 smart meters), the adaptation to diverse attack policies is not discussed.} In 5G Industrial Cyber-Physical Systems (ICPS), Li et al.\cite{LiDual} constructed a dual-network model to predict attack paths. However, their model depends on static topology and vulnerability information; when network configurations change, the dual-network model must be reconstructed and retrained.

In recent years, with the advancement of LLMs and LLM Agents\cite{zhao2023survey, achiam2023gpt, xi2025rise}, agent-based defense methods have begun to emerge. For example, IDS-Agent\cite{li2025idsagent} constructs an LLM-based agent that integrates multiple model tools and external memory retrieval to achieve explainable intrusion detection in IoT. \revise{LLM4Game\cite{peng2025llm4game} is a MARL defense method based on knowledge injection, which uses strategy vectors generated by LLMs to dynamically adjust reward functions. However, due to inherent limitations of LLMs, their in-depth application faces challenges. First, LLMs have limitations in numerical computation\cite{mirzadeh2024gsm}, as their mathematical reasoning is not based on formal logic but on abstract probabilistic matching, making it difficult to reliably handle high-dimensional, dynamic network state spaces. For instance, LLMs may struggle to accurately interpret node connectivity represented by network adjacency matrices or calculate the shortest distance from attackers to critical assets. Second, LLMs are better at macro-level planning than constrained decision-making\cite{sun2023evaluating}, making it difficult to directly generate executable defense commands. For example, even if an LLM can generate a tactical intent such as isolating the infected network segment, it may struggle to translate this into valid flow table modification.}

Currently, no framework systematically addresses these challenges. To fill this research gap, the proposed CyberOps-Bots innovatively adopts a hierarchical LLM+RL architecture, combining the semantic understanding and planning capabilities of LLMs with the atomic defense action orchestration capabilities of RL in dynamic environments. The upper-layer LLM agent is responsible for understanding the \revise{attack-defense situation}, tactical planning, and resource scheduling, while the lower-layer RL agents execute specific atomic defense actions. This approach leverages the strengths of LLMs while effectively circumventing their limitations, offering a new \revise{paradigm} for building adaptive cloud network defense systems.

To clearly demonstrate the coverage of adaptive capabilities in existing research, Table \ref{tab:works} summarizes the abilities of related works in addressing adaptability challenges A1–A4. As can be seen from the table, there is currently a lack of a complete solution capable of simultaneously meeting all four adaptability requirements mentioned above.

\section{Methodology}

\subsection{CyberOps-Bots Framework}

\begin{figure}[t]
	\centering
	\includegraphics[width=\columnwidth]{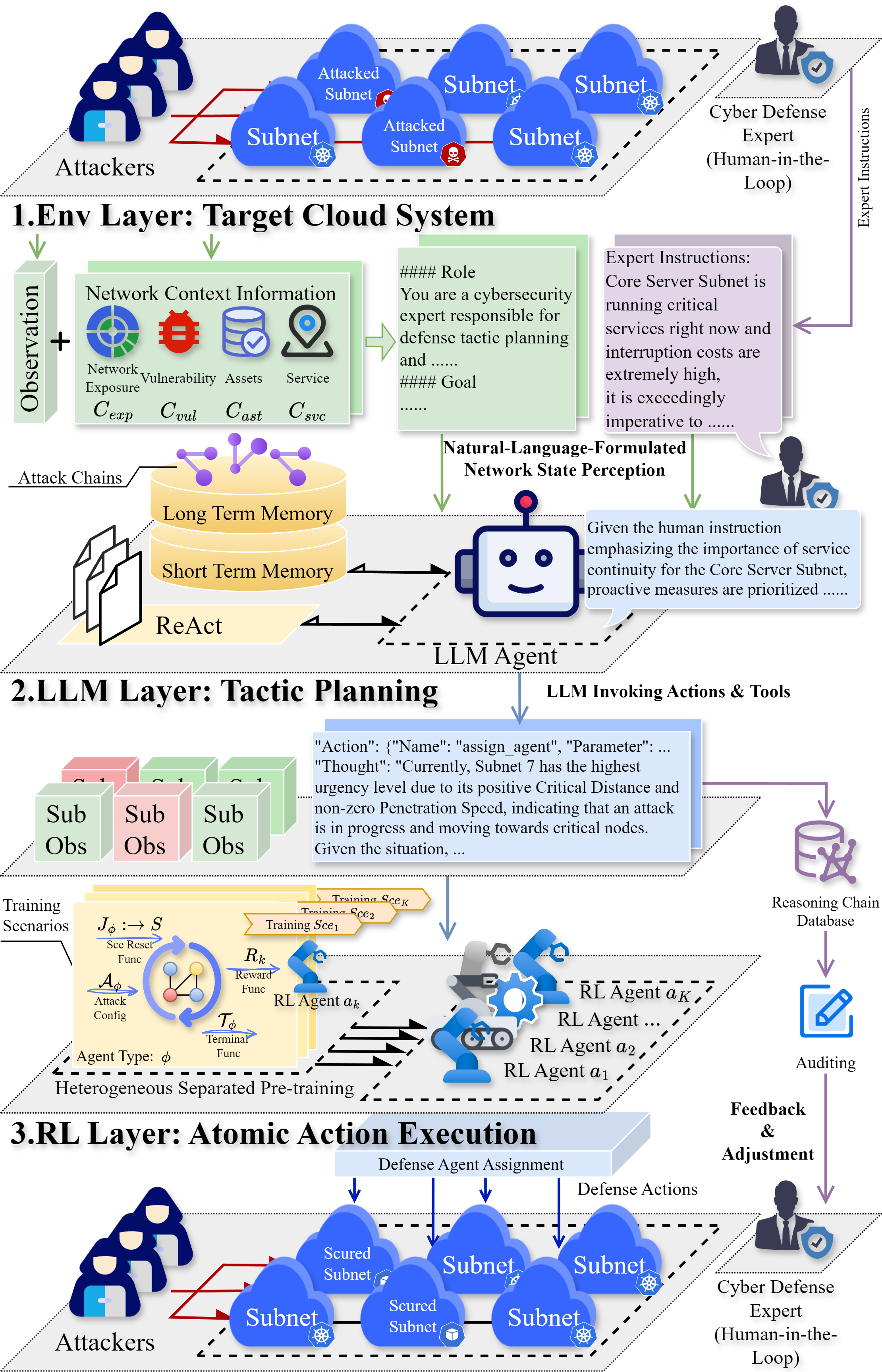} 
	\caption{The CyberOps-Bots framework architecture, comprising three coordinated layers: (i) the Env Layer simulating a dynamic adversarial cloud network; (ii) the LLM Layer with Perception, Planning, Memory, and Action modules for semantic reasoning and HITL support; and (iii) the RL Layer of pre-trained heterogeneous agents executing localized defense actions, enabling adaptive response without retraining.}
	\label{fig:framework}
\end{figure}

The overall architecture of the CyberOps-Bots framework is illustrated in Fig.\ref{fig:framework}, which consists of three core layers: the Env Layer, the LLM Layer, and the RL Layer. These correspond respectively to three stages: environmental interaction, tactical planning, and action execution.

The Env Layer simulates a dynamic adversarial scenario in a cloud environment, where attackers perform penetration and lateral movement across multiple virtual subnets, providing a real-time confrontation environment for the agent framework.

\mrevise{The upper-layer LLM agent plays four roles in our framework. First, it drives the perception module, converting high-dimensional structured network states and unstructured contextual information into unified semantic descriptions based on the NIST IPDRR framework\cite{IPDRR_nist_2025}. This perception mode ensures defense decisions do not rely on specific network topologies or scales, enabling inherent adaptation to dynamic changes in network structure (A1) and scale (A2) without retraining.
Second, the LLM executes high-level tactical planning with strong generalization. Through the ReAct paradigm, it conducts multi-step reasoning to dynamically orchestrate and schedule heterogeneous lower-layer RL agents via tool calls. This semantic-based planning capability enables flexible handling of unseen multi-stage attacks (A3), rather than merely relying on predefined strategy combinations.
Concurrently, through integration with long- and short-term memory mechanisms, the LLM maintains and analyzes attack chains across multiple time steps. It retrieves historical attack patterns via similarity-based matching to infer attack intentions and propagation trends, enabling effective response to changes in attack scales (A4) and ``reactive'' memory-based tactical planning. Traditional planners lacking semantic association and context preservation do not possess this capability.}
Finally, as an HITL interface, the LLM integrates expert knowledge and intervention via the perception module. Its auditable reasoning chains incorporate human judgment into the decision cycle, enhancing system trustworthiness in adversarial environments.

The lower-layer RL agents comprise a set of functionally heterogeneous agents. These agents are pre-trained offline in specially constructed scenarios, with each agent responsible only for its assigned segment of the network. \revise{This design avoids state space explosion due to dynamic network scale (A2).}

\subsection{Framework Formulation}
This section models the dynamics of cloud network attack and defense using a MDP and formally defines the proposed hierarchical multi-agent framework, CyberOps-Bots. The MDP process is designed in conjunction with the Yawning Titan\cite{YT} cyber attack-defense simulation platform, developed by the UK Defence Science and Technology Laboratory (Dstl). This platform has been widely adopted in academic research and practical cyber exercises, with its simulation environment validated for adversarial complexity and strategy evaluation effectiveness\cite{purves2024causally, ThompsonEntity}.

\begin{definition}
\label{def:framework}
The CyberOps-Bots framework is formally represented as a quadruple $\left(LLM, \mathcal{A}_{\text{lower}}, M, T\right)$, where:
\begin{itemize}
    \item $LLM$ denotes the upper-layer LLM agent, responsible for global tactical planning.
    \item $\mathcal{A}_{\text{lower}} = \{a_1, a_2, \ldots, a_K\}$ represents the set of lower-layer RL agents, each executing atomic defense actions within localized network regions.
    \item $M$ denotes the memory module, comprising short-term and long-term memory.
    \item $T$ denotes the toolset, which the LLM can call upon to implement tactical plans.
\end{itemize}
\end{definition}

\begin{definition}
The MDP is defined as a $(S, A, P, R, \gamma)$, where:
\begin{itemize}
    \item $S$ is the state space.
    \item $A$ is the action space.
    \item $P: S \times A \times S \rightarrow [0,1]$ is the state transition probability function.
    \item $R: S \times A \times S \rightarrow \mathbb{R}$ is the reward function.
    \item $\gamma \in (0,1)$ is the discount factor.
\end{itemize}
\end{definition}

\subsubsection{State}
This section defines the global state space, the upper-layer observation space, and the lower-layer observation space.

\begin{definition}
\label{def:state}
The global state \( s_t \in S \) at time \(t\) captures the complete security posture of the cloud network and is defined as:
\begin{equation}
	s_t = \{C, G_t, h_t(n), q_t(n), \text{vuln}_t, \text{HVN}, \text{Entry}, I_t^{\text{Human}}\}, 
\end{equation}
where:
\begin{itemize}
    \item \(C\) is the network context, which consists of semantic information about network assets, vulnerabilities, and service continuity, organized based on the NIST IPDRR framework \cite{IPDRR_nist_2025}. It provides standardized natural language input for the upper-layer LLM agent's situational awareness (see Section \ref{sec:perception}).
    \item The cloud network is represented by an undirected graph \(G_t = (V, E_t)\), whose adjacency matrix is denoted as \(Adj_t\).
    \item \(h_t(n)\) and \(q_t(n)\) represent the health status and isolation status of node \(n\), respectively.
    \item \(vuln_t(n)\) reflects the vulnerability level of node \(n\).
    \item \(HVN\) is a binary vector identifying which nodes are high-value nodes.
    \item \(Entry\) is a binary vector identifying entry nodes accessible from external attacks.
    \item \(I_t^{\text{Human}}\) represents prior knowledge or real-time intervention instructions injected by security experts via natural language at time \(t\), which is the core of the HITL mechanism. When there is no human input, \(I_t^{\text{Human}} = \varnothing\).
\end{itemize}
\end{definition}

\begin{definition}
The observation \(o_t^{LLM} \in O^{LLM}\) of the upper-layer LLM agent is a semantic perception of the global state \(s_t\), generated by the Perception module (detailed in Section \ref{sec:perception}) combining the global state and human instructions.
\end{definition}

\begin{definition}
The local observation \(o_t^k \in O_k\) of a lower-layer RL agent \(a_k\) at time \(t\) is the projection of the global state \(s_t\) onto the local network region \(G_t' \subseteq G_t\) assigned to that agent (detailed in Section \ref{sec:agent_setting}).
\end{definition}

Since the LLM agent perceives the network state in natural language, and each RL agent only perceives local observations, their observation spaces maintain structural and semantic consistency even when the network structure and scale change, thus requiring no retraining for adaptation.

\subsubsection{Action}
For the defender, the proposed framework defines two distinct action spaces: a high-level tactical action space $A^{\text{LLM}}$ for the upper-layer LLM agent and a low-level atomic action space $A^{\text{lower}}$ for the lower-layer RL agents.

\begin{definition}
The atomic action space $A^{\text{lower}}$ consists of fundamental defensive operations applicable to cloud network nodes. An atomic action $act_k \in A^{\text{lower}}$ is defined as a tuple $(operation, target)$, where $operation$ denotes the type of defensive action (e.g., resetting a container image, patching a node vulnerability, isolating a compromised container, or restoring a node's service), and $target$ specifies the node $n \in V$ to which the action is applied.
\end{definition}

\begin{definition}
The tactical action space $A^{LLM}$ for the upper-layer LLM agent comprises the set of tactical instructions it can execute. This includes both directly invoking atomic actions and calling tools, formally represented as $A^{L L M}=\left\{A^{\text{lower}},\mathcal{T}\right\}$.
\end{definition}

For the attacker, actions are modeled based on penetration capabilities. Attackers can initiate attacks from any compromised node or entry node towards adjacent nodes. The success probability $P_{\text{attack}}\left(s_t, R S\right)$ of an attack on a visible target node $n_{\text{target}}$ depends on the target node's vulnerability and the attacker's skill level $R S\in[0,1]$ \cite{purves2024causally}, calculated as:
\begin{equation}
	P_{\text{attack}}\left(s_t, R\,S\right)=\min\left(\frac{R\,S^2}{R\,S+\left(1-\mathrm{vuln}_t\left(n_{\mathrm{target}}\right)\right)}, 1\right).
\end{equation}
If the attack is successful, the health state of the target node is updated to Compromised: $h_{t+1}\left(n_{\text{target}}\right)=Compromised$.

\subsubsection{Global Reward Function}

This section designs the global reward function.

\begin{definition}
To ensure the reward design aligns with the practical requirements of cloud network defense, the reward objectives are established with reference to the three aspects defined in the ISO/IEC 27001:2022 standard \cite{isoiec_2022}: asset management, network security, and defense continuity. The global reward function \(R\) is formulated as:

\begin{equation}
	R\left(s_t, a_t, s_{t+1}\right) = R_{\text{asset}}\left(s_{t+1}\right) + R_{\text{security}}\left(s_t, s_{t+1}\right) + R_{\text{cost}}\left(a_t\right), 
\end{equation}

where:

\begin{enumerate}[label=\alph*)]
    \item \textbf{Asset Protection Reward (\(R_{\text{asset}}\))}:
    
    \begin{equation}
    	R_{\text{asset}}\left(s_{t+1}\right) = -\lambda_{hva} \cdot \sum_{n \in V} HVN_t(n) \cdot h_{t+1}(n).
    \end{equation}
    
    This component penalizes the compromise of high-value nodes (HVNs), aiming to prevent critical data leakage or tamper. A significant penalty is incurred when an HVN is compromised.

    \item \textbf{Network Security Reward (\(R_{\text{security}}\))}: This term penalizes all nodes in an abnormal state (e.g., compromised or isolated) within the network. It comprehensively considers both the quantity and the change in abnormal nodes.
    \begin{equation}
    	R_{\text{security}}\left(s_t, s_{t+1}\right) = -\left(\alpha \cdot N_{t+1}^{\text{impact}} + \beta \cdot \Delta N_t^{\text{impact}}\right).
    \end{equation}
    Here, \(N_{t+1}^{\text{impact}}\) denotes the number of abnormal nodes at time \(t+1\), and \(\Delta N_{t}^{\text{impact}}\) represents the number of newly added abnormal nodes from time \(t\) to \(t+1\). The coefficients \(\alpha\) and \(\beta\) are weighting factors.

    \item \textbf{Action Cost Reward (\(R_{\text{cost}}\))}:
    \begin{equation}
    	R_{\text{cost}}\left(a_t\right) = -Cost(operation).
    \end{equation}
\end{enumerate}
\end{definition}

\subsection{Lower-Layer RL Agents}

The lower layer of the framework comprises multiple types of specially trained RL agents, denoted as the set $\mathcal{A}_{\text{lower}} = \{a_1, a_2, \ldots, a_{K}\}$ in Definition \ref{def:framework}. \mrevise{These agents possess diverse defense objectives and specialize in different action types. Through the tactical planning and orchestration by the upper-layer LLM, they are combined to form varied defense strategies, thereby enabling the framework to adapt to different attack policies (A3).}

\subsubsection{Agent Setting}

\label{sec:agent_setting}

For a lower-layer RL agent $a_k$ assigned to a subnet $G_{t}^{\prime} \subseteq G_{t}$ at time step $t$, its decision-making is based on local observation information within that network segment. This design prevents the state space from exploding as the network scale increases.

\begin{definition}
The sub-observation space and sub-action space for an agent are subsets of their global counterparts, formally defined as follows:
\begin{enumerate}[label=\alph*)]
    \item \textbf{Sub-observation space $O_k$}: Consists of the local network state variables that RL agent $a_k$ focuses on, such as node health status, vulnerability level, isolation status, network topology connections, and high-value node distribution. Its specific composition depends on the agent's defensive expertise and task objectives.
    \item \textbf{Sub-action space $A_k$}: Contains the types of defensive actions executable by agent $a_k$ and their scope of application, typically represented as $A_k = \{(action, n) \mid action \in O_k, n \in G_t^{\prime}\}$, where $O_k \subseteq operation$ is the set of defensive actions that the agent can perform.
\end{enumerate}
The determination of the above observation and action spaces depends on the function of the corresponding agent. For example, consider a \textit{Block Agent}. This agent is responsible for precisely blocking attack paths by executing the \textit{Isolate} action when attackers approach high-value nodes. Therefore, this agent needs to comprehensively analyze the network topology, node health status, and high-value node distribution to accurately determine isolation targets. Its sub-observation space is defined as $O_{\text{iso}} = \{Adj_{t}, h_{t}(n), HVN \mid \forall n \in G_t^{\prime}\}$, and its sub-action space is defined as $A_{iso} = \{(Isolate, n) \mid \forall n \in G_t^{\prime}\}$.
\end{definition}

\subsubsection{Heterogeneous Separated Pre-training}
\label{sec:hsp}

We introduce a heterogeneous separated pre-training framework for lower-layer RL agents, integrating reward function design and scenario construction. \mrevise{This guarantees stable training of specialized agents with differentiated defensive skills, providing the upper-layer tactical planner with diverse defenses to address dynamic attack policies (A3).}

\begin{definition}
To guide the lower-layer RL agents in learning and reinforcing their specific decision-making styles, we design specialized reward functions for each type of agent, forming a reward function set $\mathcal{R}_{\text{lower}} = \{R_1, R_2, \ldots, R_K\}$. These functions are built upon the global reward function $R(s_t, a_t, s_{t+1})$, but place greater emphasis on the objectives most relevant to their respective subtasks, thereby training functionally heterogeneous expert agents.
\end{definition}

This independent training design addresses the issue of training instability in Hierarchical Multi-Agent Reinforcement Learning (HMARL) \cite{singh2024hierarchical}. It specifically tackles the instability caused by the interdependencies that exist between policy hierarchies, among agents, and across different subtasks during the HMARL training process.

\begin{definition}
A training scenario for an agent of type $\phi$ is formally defined as a triple $(\mathcal{J}_{\phi}, \mathcal{A}_{\phi}, \mathcal{T}_{\phi})$, where:
\begin{itemize}
    \item $\mathcal{J}_{\phi}: \rightarrow S$ is the scenario reset function, which generates an initial state $s_0$ aligned with the learning objectives of agent type $\phi$;
    \item $\mathcal{A}_{\phi}$ is the attack configuration, defining the number of attackers and their behavior patterns within the scenario;
    \item $\mathcal{T}_{\phi}: S \rightarrow \{0,1\}$ is the termination condition function.
\end{itemize}
\end{definition}

Through the construction of these training scenarios and the independent training of different RL agents, the method resolves the agent training dependency problem inherent in HMARL \cite{singh2024hierarchical}. It enables each agent type to learn its policy in an environment highly tailored to its responsibilities, ensuring both training stability and efficiency.

\begin{algorithm}
\caption{Heterogeneous Separated Pre-training.}
\label{alg:training}
\begin{algorithmic}[1]
\STATE \textbf{Input:} Agent type $i$, training episode length $T$,
\STATE \textbf{Output:} Trained policy network parameter $\theta_{i}$
\FOR{$e$ in $E$}
    \STATE Initialize scenario $S_{i}^{e}$
    \STATE Reset environment with $S_{i}^{e}$:  $s_{0} \sim \mathcal{R}(S_{i}^{e})$
    \FOR{$t$ in $T$}
        \STATE Observe: $o_t^i$
        \STATE Select action with policy $\pi_{\theta_i}$: $a_t^i \sim \pi_{\theta_i}(o_t^i)$
        \STATE Execute $a_t^i$ and attack according to $A(S_{i}^{e})$: $s_{t+1}, r_t^i \sim Env(s_t, a_t^i)$
        \STATE Calculate reward $R_i(s_t, a_t^i, s_{t+1})$
        \STATE Append $(s_t, a_t^i, r_t^i, s_{t+1})$ to buffer $D$
        \IF{learn}
            \STATE Sample a training batch from $D$
            \STATE Update $\theta_i$ by minimizing loss
        \ENDIF
    \ENDFOR
\ENDFOR
\STATE Return $\theta_{i}$
\end{algorithmic}
\end{algorithm}

The heterogeneous separated pre-training method is shown in Algorithm \ref{alg:training}, which can be adapted to any single-agent RL algorithm for agent parameter updates.

\subsection{Upper-Layer LLM Agent}

The upper layer of the proposed hierarchical framework is an LLM agent. \revise{Serving as the global decision-maker, its core function is to integrate the execution capabilities of RL agents with high-level semantic planning to address adaptability challenges in cloud networks.}

\revise{The LLM agent architecture comprises four core components\cite{xi2025rise}: Planning, Perception, Memory, and Action and Tools. Specifically, the agent transforms state information into natural language via the Perception module, utilizes the Planning module for multi-step reasoning and tactical formulation, maintains attack chain context through the Memory module, and translates decisions into concrete defensive actions or scheduling instructions for lower-layer RL agents via the Action and Tools module.}

The following subsections detail each module.

\subsubsection{Planning}

To achieve adaptive planning, the upper-layer LLM agent employs the ReAct\cite{yao2022react, ShinnAdvances}. ReAct is a prompting technique that synergizes reasoning and acting. By guiding the LLM to explicitly generate reasoning traces and subsequent actions, it significantly enhances the reliability of decision-making in complex tasks.

At each time step \(t\), the planning process of the LLM agent is as follows:
\begin{enumerate}[label=\alph*)]
    \item \textbf{Observation}: The LLM receives network situational information from the Perception module, the Long-Term Memory \(LTM_{t}\) and Short-Term Memory \(STM_{t}\) stored in the Memory module, as well as a list of available tools and actions (including their functions, parameters, etc.).
    \item \textbf{Reasoning}: Based on the input observation information \(O_{t}^{LLM}\), the LLM generates a reasoning text. By analyzing the situation and objectives, it formulates a strategy and concludes the reasoning process.
    \item \textbf{Acting}: Following the reasoning, the LLM outputs the corresponding action and its parameters, denoted as \(\left(a_{t}^{LLM}, para_{t}\right)\). After the environment executes the action, the state information is updated, new memories are generated, and the cycle repeats.
\end{enumerate}

The reasoning chain generated by the LLM within the ReAct loop is saved as logs, providing an interpretable audit trail for expert review.

\subsubsection{Perception}

\label{sec:perception}

The perception module acts as the information gateway for the upper-layer LLM agent, converting high-dimensional, structured network state data into natural language. This abstraction decouples defense decisions from specific topologies and scales, ensuring adaptability to structural (A1) and scalar (A2) changes. \revise{Crucially, this semantic unification serves as the cornerstone of the framework's robustness, empowering the agent to generalize to unseen network configurations seamlessly without retraining.} Additionally, by integrating expert knowledge and human instructions, the module enables a HITL interaction mechanism.

Inputs to this module include the network context $C$, expert instructions $I_t^{Human}$, and calculated state metrics. These are aggregated via a prompt template to generate the LLM's observation $O_t^{LLM}$.

Organized by the NIST IPDRR cybersecurity framework \cite{IPDRR_nist_2025}, the module processes the following context and metrics:

a) \textbf{Identify}: Identifies potential threat surfaces and propagation risks.
\begin{itemize}
    \item Subnet Exposure Surface $C_{exp}$: Defines connection policies and boundaries between subnets and untrusted networks, identifying the attack surface.
    \item Entry Node Count: Quantifies the initial size of the subnet's attack surface.
\end{itemize}

b) \textbf{Protect}: Evaluates asset vulnerabilities and the effectiveness of existing defenses.
\begin{itemize}
    \item Subnet Vulnerability Profile $C_{vul}$: details device types, firmware, CVEs, and patch status, outlining inherent asset vulnerabilities.
    \item Average Vulnerability: Quantifies the subnet's overall vulnerability level.
    \item Compromised Node Count: Measures the extent of damage and current defense failure.
\end{itemize}

c) \textbf{Detect}: Analyzes attack trends, strategies, and patterns.
\begin{itemize}
    \item Attack Chain Concentration: Measures strategy concentration via information entropy to identify attack patterns:
    \begin{equation}
    	H_{\text{attack}}(t)=-\sum_{i=1}^k p_i\log\left(p_i\right), 
    \end{equation}
    where $p_i$ is the proportion of attacks on the $i$-th subnet over time $\Delta T$. Lower entropy indicates a more deterministic strategy.
    \item Attack Frequency: The average attacks per subnet $G'$ over $\Delta T$, reflecting attacker activity and focus.
\end{itemize}

d) \textbf{Respond}: Details asset urgency to guide response prioritization.
\begin{itemize}
    \item HVN Count: The number of HVNs, indicating subnet criticality.
    \item Critical Distance: The shortest path from compromised nodes to HVNs, reflecting immediate threat urgency. It is defined as:
\begin{equation}
	\begin{aligned}
		D_t^{\text{critical}}(G') = \min \{ & \text{DisToHVN}(n_c) \mid \\
		& n_c \in G', h_t(n_c)=\text{Compromised} \}.
	\end{aligned}
\end{equation}
    \item Penetration Speed: The rate of change in Critical Distance, $\Delta D_{t}^{\text{critical}}(G')=D_{t-1}^{\text{critical}}(G')-D_{t}^{\text{critical}}(G')$. Positive values indicate attack progression; negative values imply effective defense.
\end{itemize}

e) \textbf{Recover}: Plans recovery actions for business continuity.
\begin{itemize}
    \item Subnet Service Continuity $C_{svc}$: Includes RTO and RPO policies, setting business recovery priorities.
    \item Isolated Node Count: Indicates the severity of service disruption.
    \item Connectivity: Measures the impact of isolation on availability. For a subnet $G_{t}^{\prime}=(V^{\prime}, E^{\prime})$, connectivity is the ratio of actual to maximum theoretical edges:
    \begin{equation}
    	\operatorname{Conn}\left(G_t^{\prime}\right)=\frac{\left|E_t^{\prime}\right|}{\left|V^{\prime}\right|\times\left(\left|V^{\prime}\right|-1\right)/ 2}.
    \end{equation}
\end{itemize}

\revise{The above contextual information and key metrics are concatenated via prompts to form the natural language observation information \(O_t^{LLM}\). To ensure the reproducibility of our study, the complete collection of prompt templates utilized in the perception module is disclosed in supplemental files.}

\subsubsection{Memory}
\label{sec:memory}

To cope with the dynamic changes in attack strategies and scales, the LLM agent requires the capability to remember and analyze past attack events, enabling long-term planning and multi-attack-chain analysis across time steps. This memory-driven adaptability is essential for the framework's robustness, as it allows the agent to identify and counter evolving threats in real-time without requiring retraining on new attack patterns. In LLM-based agent frameworks, Memory refers to mechanisms maintaining historical interactions to support current decisions. This paper designs two memory mechanisms: Long-Term Memory and Short-Term Memory, detailed as follows:

\begin{enumerate}[label=\alph*)]
    \item \textbf{STM}: STM maintains immediate context within the most recent decision cycle, containing the agent's previous action and environmental feedback. Specifically, STM at time $t$ is a triple:
    \begin{equation}
    	STM_t=\left(a_{t-1}^{LLM}, O_{t-1}^{LLM}, s_t\right).
    \end{equation}
    STM ensures coherent reasoning trajectory generation, allowing the LLM to understand environmental changes caused by its previous actions. This supports continuous strategy optimization and avoids myopic reasoning or repeated decisions due to information fragmentation.

    \item \textbf{LTM}: \revise{LTM acts as an external knowledge base for persistently storing and analyzing attack chains. To ensure scalability and prevent context overflow during long-term campaigns, LTM employs a sliding temporal window $W$ to retain only active trajectories. At time $t$, $LTM_t$ is defined as the set of attack chains occurring within the interval $[t-W, t]$:}
    \begin{equation}
    	LTM_t = \{ AC_1, AC_2, \dots, AC_k \mid \forall (n, \tau) \in AC_i, \tau \in [t-W, t] \}.
    \end{equation}
    \revise{Each attack chain $AC_i$ is represented as a sequence of compromised events within the subnet $G_i^{\prime}$:}
    \begin{equation}
    	AC_i = \left( G_i^{\prime}, \left[ (n_{i, 1}^{\text{target}}, t_{i, 1}), (n_{i, 2}^{\text{target}}, t_{i, 2}), \dots \right] \right).
    \end{equation}
    \revise{Here, each tuple $(n^{\text{target}}, \tau)$ records a target node $n^{\text{target}}$ compromised at time $\tau$. By discarding historical data outside the window $W$, this mechanism ensures that the memory size remains bounded and the inference efficiency is preserved regardless of the attack duration.}
\end{enumerate}

\begin{algorithm}
\caption{Reactive Retrieval and Analysis of LTM}
\label{alg:memory}
\begin{algorithmic}[1]
\STATE \textbf{Input:} Attack chains $L_c$, Long-term Memory $G_p$, Critical distance threshold $\theta$, Top-k similarity threshold $\delta$
\STATE \textbf{Output:} Predicted attack chain $L_p$
\IF{CriticalDistance $< \theta$}
    \STATE $L_p \leftarrow \emptyset$
    \STATE Return $L_p$
\ENDIF
\FOR{$L_i$ where $L_i \in G_p$}
    \STATE Match score $M_s \leftarrow \text{similarity}(L_c, L_i)$
\ENDFOR
\STATE Rank $L_i$ by $M_s$ in descending order and select the Top-k chains
\STATE Calculate $L_p \leftarrow \text{most\_likely\_attack\_chain}(\text{Top-k})$
\STATE Select $L_p$
\STATE Append $L_p$
\STATE Integrate $L_p$ into LLM observation $O$ for tactical planning
\end{algorithmic}
\end{algorithm}

\revise{LTM retrieval is reactive. When the Critical Distance of a subnet falls below a threshold $\theta_{\text{critical}}$, the mechanism retrieves and analyzes Top-k similar chains from the memory bank, integrating the most probable attack chain into the LLM's observations.} The specific retrieval and analysis algorithm is provided in Algorithm \ref{alg:memory}.

\subsubsection{Action and Tools}

The upper-layer LLM agent translates defense decisions into specific defense instructions and actions through the Action and Tools module. This module provides the following two types of tools:

\begin{enumerate}[label=\alph*)]
    \item Action Execution Functions: These functions enable the LLM agent to directly execute defense actions on network nodes. An action execution function is defined as \texttt{ExecuteAction(operation, n)}, where the input consists of the action type and the target node.

    \item Lower-layer Agent Assignment Tools: These tools are used to dispatch lower-layer RL agents to specific subnets. A tool can be defined as \texttt{AssignAgent($\phi$, $G^{\prime}$)}, where $\phi$ represents the type of RL agent and $G^{\prime}$ denotes the target subnet. Within any single time step $t$, the usage of this tool by the LLM agent is constrained by the total global agent count.
\end{enumerate}

Through the design of the tools, the proposed framework achieves flexible integration of tactical planning and defense action execution. This design allows the upper-layer LLM to conduct tactical planning via the ReAct paradigm and orchestrate functionally heterogeneous lower-layer RL agents according to the real-time offensive and defensive situation. As a result, the framework combines the LLM's capabilities in planning and generalization with the specialized execution strengths of RL agents, enabling the dynamic generation and implementation of customized defense solutions tailored to specific attack strategies and stages.

\section{Experiments}

\subsection{Experimental Design}

To systematically evaluate the effectiveness of the proposed CyberOps-Bots framework, the following research questions (RQs) are designed:

\begin{itemize}
    \item \textbf{RQ1:} How robust is the CyberOps-Bots framework when facing dynamic changes in cloud network structure (A1)?
    \item \textbf{RQ2:} How robust is the CyberOps-Bots framework when facing dynamic changes in cloud node scale (A2)?
    \item \textbf{RQ3:} Is the CyberOps-Bots framework robust against unseen and diverse cloud \mrevise{attack policies} (A3)?
    \item \textbf{RQ4:} Can the CyberOps-Bots framework effectively handle changes in attack scale caused by an increasing number of concurrent attackers (A4)?
    \item \textbf{RQ5:} Can the CyberOps-Bots framework balance defense effectiveness and network availability to enhance cyber resilience?
    \item \textbf{RQ6:} Can the CyberOps-Bots framework improve long-term network resilience through proactive hardening?
    \item \textbf{RQ7:} How is the decision-making efficiency and reliability of the CyberOps-Bots framework?
    \item \textbf{RQ8:} Can the CyberOps-Bots framework effectively support HITL?
\end{itemize}

\begin{figure}[tbp]
	\centering
	\includegraphics[width=\columnwidth]{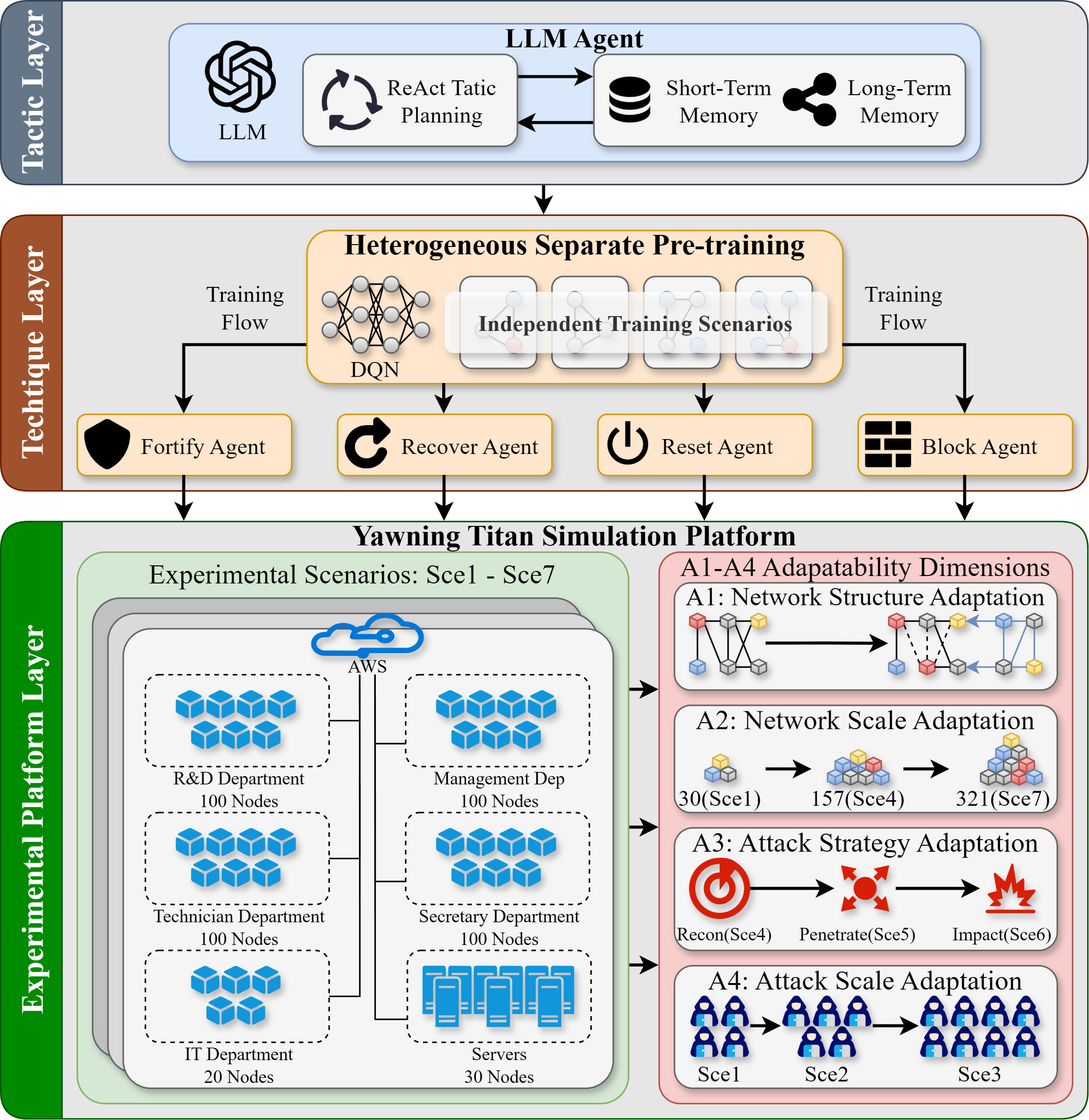} 
	\caption{Experimental setup for evaluating the adaptability of CyberOps-Bots and baseline algorithms.}
	\label{fig:experiment_senario}
\end{figure}

\mrevise{This study designs experimental scenarios based on the AWS enterprise cloud dataset\cite{dataset}.} The scenario configurations are dynamically switched to validate A1-A4, as shown in Fig.\ref{fig:experiment_senario}. The network scale, structure, exposure surface, asset distribution, vulnerability configuration, and service continuity requirements of the scenarios are designed based on the metadata and attack log data from this dataset. The dataset is also used to summarize the network context $C$ for the LLM, as described in Section \ref{sec:perception}. Specifically:

\begin{itemize}
\item \textbf{Scale and Structure:} Mirroring the dataset's infrastructure, the experimental network comprises 450 nodes divided into 6 subnets (5 departments and a server room). The node count in each subnet corresponds to the actual device distribution of the respective departments.
\item \textbf{Network Exposure:} Gateway nodes and inter-subnet topology are configured according to the dataset's connection and isolation policies, aligning east-west traffic with real segmentation requirements. These boundary definitions and connection policies constitute the context $C_{\text{exp}}$.
\item \textbf{Assets:} HVNs, such as Web and Database servers, are designated based on departmental business functions. These business descriptions and HVN distributions are summarized as the asset context $C_{\text{ast}}$.
\item \textbf{Vulnerability:} Nodes are assigned vulnerability values based on the dataset's specific attack types and CVE records. This information is summarized as the vulnerability context $C_{\text{vul}}$.
\item \textbf{Service Continuity:} Operational policies for Recovery Time/Point Objectives (RTO/RPO) are derived from the dataset's business metadata and service continuity agreements, summarized as the service continuity context $C_{\text{svc}}$.
\end{itemize}

\begin{table*}[htbp]
\footnotesize 
\caption{Subnet Configurations of the Experimental Cloud Network.}
\label{tab:subnet_config}
\centering

\begin{tabularx}{\textwidth}{T T L L T}
\toprule
Subnet Name & Node Scale & Service & Key Assets & Entry nodes \\ 
\midrule 

Dep1 & 100 & Research and Development department. This subnet generates internal R\&D traffic and accesses internal servers. Hosts various client workstations used by researchers. & - R\&D Workstation \newline - Internal Code Repository Server & 5 \\ 

Dep2 & 100 & Management Department. Handles administrative functions and internal communications. Traffic may include access to financial records and management systems. & - Management Workstation \newline - Internal File Share Server & 6 \\ 

Dep3 & 100 & Technician Department. Responsible for technical support and infrastructure monitoring. May have higher network privileges and generate different traffic patterns. & - Technician Workstation \newline - Network Monitoring Server \newline - Patch Management Server & 4 \\ 

Dep4 & 100 & Secretary and Operation Department. Handles day-to-day operational tasks, external communications, and customer-facing activities. & - Customer Relations Management (CRM) Server \newline - Email Server & 5 \\ 

Dep5 & 20 & IT Department. This subnet manages IT infrastructure and has access to core network services. & - IT Administrator Workstation \newline - DNS Server & 3 \\ 

Servers & 30 & Central server room hosting critical services for the entire organization. Includes web servers, database servers, and other infrastructure servers. & - Apache Web Server \newline - MySQL Database Server \newline - Windows Server 2012/2016 (For various enterprise services) & 2 \\ 
\bottomrule
\end{tabularx}
\end{table*}

The complete experimental subnet configurations are shown in Table \ref{tab:subnet_config}.

\revise{To account for variations in attack strategies and phases\cite{singh2024hierarchical}, this paper designs three types of attack policies based on the scenario envisioned in the introduction.} These correspond to different ATT\&CK Tactics\cite{ATTCK}. The first is \textbf{Recon}, aligning with the Reconnaissance and Discovery tactics. Its logic involves attackers randomly probing reachable nodes to assess the attack surface. The second is \textbf{Penetrate}, corresponding to the Initial Access and Lateral Movement tactics, where attackers prioritize targeting nodes with higher vulnerability values to achieve infiltration and spread within the network. The third is \textbf{Impact}, associated with the Collection and Impact tactics, where attackers plan optimal paths based on network information to directly target high-value nodes, aiming to cause service disruption or data theft.

\begin{table*}
	\footnotesize 
	\caption{Experimental scenarios.}
	\label{tab:senarios}
	\begin{tabularx}{\textwidth}{cc Y cc c}
		\toprule 
		Scenario & Node & \multirow{2}{*}{Subnets in Use} & \multicolumn{2}{c}{Attack   Setting} & \multirow{2}{*}{Validation Objective} \\ \cmidrule(lr){4-5} 
		Name & Scale &  & Attacker Scale & Attack Policy &  \\ \midrule 
		
		Sce1 & \multirow{3}{*}{30} & \multirow{3}{*}{Servers} & 6 & \multirow{3}{*}{recon} & A1, A2, A4 \\
		Sce2 &  &  & 7 &  & A4 \\
		Sce3 &  &  & 8 &  & A4 \\ \midrule 
		
		Sce4 & \multirow{3}{*}{150} & \multirow{3}{*}{Servers, Dep1, Dep2} & \multirow{3}{*}{6} & recon & A2, A3 \\ 
		Sce5 &  &  &  & penetrate & A3 \\
		Sce6 &  &  &  & impact & A3 \\ \midrule 
		
		Sce7 & 450 & Servers, Dep1, Dep2, Dep3, Dep4, Dep5 & 6 & recon & A2 \\ \bottomrule 
	\end{tabularx}
\end{table*}

Different scenarios are switched during the experiments to validate the adaptability of various methods. The LLM used in this paper is Qwen3-8B\cite{qwen}. The experimental scenarios are built upon the cyber defense simulation environment Yawning Titan\cite{YT}, and the baseline algorithms are implemented based on XuanCe\cite{xuance}. \revise{The simulation environment can serve as a benchmark for trajectory comparison. The framework calculates expected environment state based on the RL agents’ policies. Comparing this state with the actual state from human intervention quantifies the discrepancy between actual tactical commands and the RL policy. These deviations are recorded in auditable decision logs, providing a data foundation for analyzing human expertise and iteratively refining the collaborative strategy.} The configured experimental scenarios are shown in Table \ref{tab:senarios}. 

\revise{Following ISO/IEC 27001:2022 security domains (Protect, Resilience, Defend) \cite{isoiec_2022}, this paper develops four heterogeneous lower-layer RL agents. The Fortify Agent hardens networks via Patch actions (Protect) to minimize attack surfaces. The Recover Agent uses Recover actions (Resilience) to restore services post-isolation, ensuring continuity. The Purge Agent integrates Image Reset and Patch actions (Defend) to eliminate threats and rebuild nodes. The Block Agent employs Isolate operations (Defend) to sacrifice connectivity for asset protection and attack containment.}

\begin{table}[htbp]
\centering
\caption{Configuration for DQN training of lower-layer RL agents.}
\label{tab:dqn_params}
\begin{tabular}{lc} 
\toprule
Parameter & Value \\
\midrule
Learning rate & 0.01 \\
Batch size & 256 \\
Discount factor $\gamma$ & 0.8 \\
Buffer size & 100,000 \\
Optimizer & Adam \\
Episodes & 10,000 \\
Activation function & ReLu \\
Target update internal & 1000 \\
$\epsilon$-greedy exploration factor & $\epsilon_{start}=0.6,\epsilon_{end}=0,p=2000$ \\ 
CPU & 16 $\times$ AMD EPYC 7453 \\
GPU & NVIDIA RTX4090 \\
Hard disk & 700GB \\
Memory & 56GB \\
Operating System & Ubuntu 20.04.1 \\
\bottomrule
\end{tabular}
\end{table}

This paper employs the DQN algorithm to train the RL agents. \mrevise{Notably, our hierarchical framework is modular and algorithm-agnostic, allowing for the seamless replacement of DQN with other RL algorithms in future usage.} The DQN network used is a two-layer fully connected neural network, with each hidden layer containing 64 nodes. Other experimental settings are detailed in Table \ref{tab:dqn_params}.

\revise{The selected baseline algorithms include IPPO\cite{IPPO}, IQL~\cite{IQL}, MAPPO~\cite{IPPO}, QMIX~\cite{QMIX}, and VDN~\cite{VDN}.}

\subsection{RQ1: Robustness against Dynamic Network Structure}

\mrevise{This experiment validates the adaptability of the proposed framework in scenarios with highly dynamic changes in network structure.}

\begin{table}[htbp]
	\centering
	\caption{Experimental Results Under Scenario 1.}
	\label{tab:rq1}
	\scriptsize
	\begin{tabularx}{\columnwidth}{ABBBBBB}
		\toprule
		Metrics & \textbf{Ours} & IPPO & IQL & MAPPO & QMIX & VDN \\
		\midrule
		Mean Reward & \textbf{-0.67} & -3.44 & -3.64 & -3.17 & -3.65 & -3.68 \\
		Reward Coefficient Variations & \textbf{0.011} & 0.059 & 0.037 & 0.057 & 0.029 & 0.028 \\
		Mean Healthy Ratio & \textbf{0.91} & 0.84 & 0.59 & 0.72 & 0.60 & 0.61 \\
		Mean Episode Length & \textbf{97.3} & 44.1 & 23.9 & 41.2 & 20.6 & 19.9 \\
		\bottomrule
	\end{tabularx}
\end{table}

\mrevise{The experiment is conducted based on Sce1. During the experiment, at random intervals, configurations including entry nodes, HVNs, and node vulnerabilities are randomly adjusted, while some nodes are randomly isolated, thereby creating a network scenario characterized by high uncertainty and dynamic changes.} Algorithm performance is evaluated using the mean reward, the coefficient of variation of the reward, the mean healthy node ratio, and the mean episode length. Here, the coefficient of variation of the reward is the ratio of the standard deviation to the mean of the average reward within 30 time steps after a dynamic change in the network, reflecting its performance fluctuation when the network structure changes dynamically. The healthy ratio refers to the proportion of nodes that are neither compromised nor isolated. The experimental results are shown in Table \ref{tab:rq1}.

Our method significantly outperforms the baseline algorithms in terms of mean reward, mean healthy node ratio, and mean episode length. \mrevise{The significantly higher mean healthy node ratio demonstrates the framework's resilience, as it successfully maintains a larger proportion of available services despite the continuous structural disruptions (A1).}

\subsection{RQ2: Robustness against Dynamic Network Scale}

To validate the adaptability of the proposed method to changes in cloud network structure and scale, this section configures experimental scenarios that dynamically transition from Sce1 to Sce4 and Sce7, thereby increasing the network node scale from 30 to 450. The adaptability to dynamic changes in network structure and scale is evaluated by comparing the average cumulative reward and jumpstart performance\cite{Zhou2025Mind, taylor2009transfer} of different algorithms. Here, jumpstart, calculated as defined in\cite{Zhou2025Mind}, refers to the average cumulative reward obtained by an agent during the first 10 episodes after training begins in a new environment. This metric reflects the algorithm's early performance during network dynamic changes, indicating its capability for rapid adaptation to dynamically changing environments.

\begin{figure}[htbp]
	\centering
	\includegraphics[width=\columnwidth]{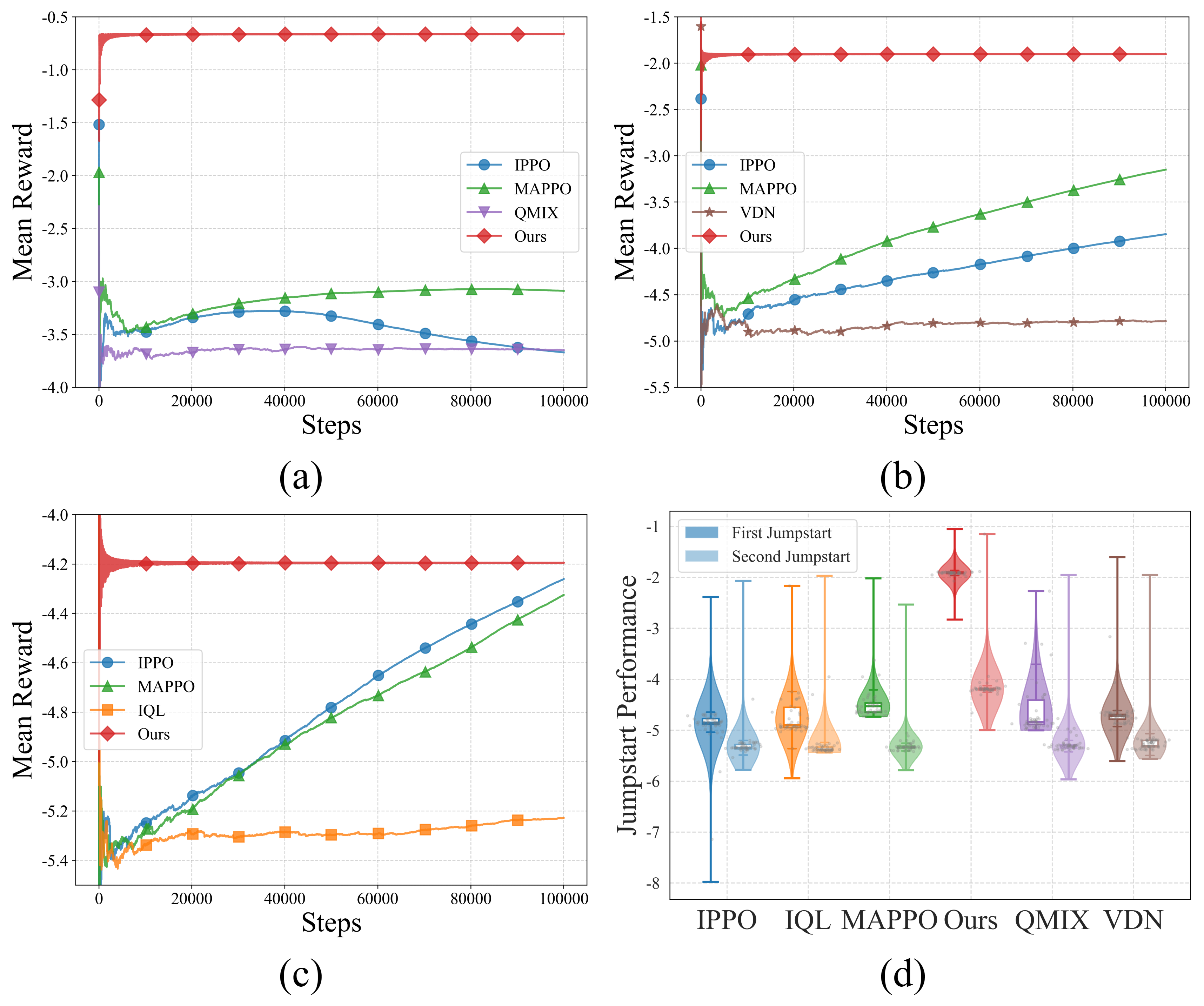} 
	\caption{Figure (a-c) present the experimental results when the test scenario is dynamically switched from Sce1 (a) to Sce4 (b) and Sce7 (c), showing the variation of average cumulative rewards over decision steps. Figure (d) illustrates the jumpstart performance of each algorithm during scenario switching, highlighting their ability to adapt quickly to new environments.}
	\label{fig:scale_line}
\end{figure}

\mrevise{As shown in Fig.\ref{fig:scale_line}(a)–(c)}, during the dynamic transition from Sce1 to Sce4 and then to Sce7, the average cumulative reward of our method is significantly higher than those of baseline algorithms. Furthermore, the reward curves converge quickly after each scenario switch and remain stable after 10,000 steps. Particularly in the Sce7 scenario (Fig.\ref{fig:scale_line}(c)), \mrevise{where the node scale expands from 30 to 450, although the reward values of IPPO and MAPPO gradually approach our method after 80,000 steps, their convergence is slow.}

\mrevise{Fig.\ref{fig:scale_line}(d) shows the jumpstart performance of each algorithm during the two dynamic transitions, reflecting their ability to quickly leverage prior experience to adapt to the cloud network and initialize strategies.} Our method achieves the best jumpstart performance, with its values being concentrated and overall higher than those of the baseline algorithms. This indicates that the method can rapidly respond to changes in network structure and scale, maintaining high defense effectiveness without retraining. \revise{This is because the upper-layer LLM agent perceives the network situation via natural language descriptions (Section \ref{sec:perception}), mapping diverse network states into a unified IPDRR-based semantic space, while the lower-layer RL agents only need to focus on local network states. Consequently, when the network scale expands, the CyberOps-Bots framework effectively avoids the problem of state space explosion, achieving seamless adaptation to the scaling of the cloud network (A2).}

\subsection{RQ3: Robustness against Evolving Attack Policy}

\mrevise{To validate the adaptability of the proposed method to different attack policies/phases, this section configures the experimental scenarios to dynamically switch from Sce4 to Sce5 and Sce6, thereby changing the attack policy from Recon to Penetrate and Impact.}

\begin{figure}[htbp]
	\centering
	\includegraphics[width=\columnwidth]{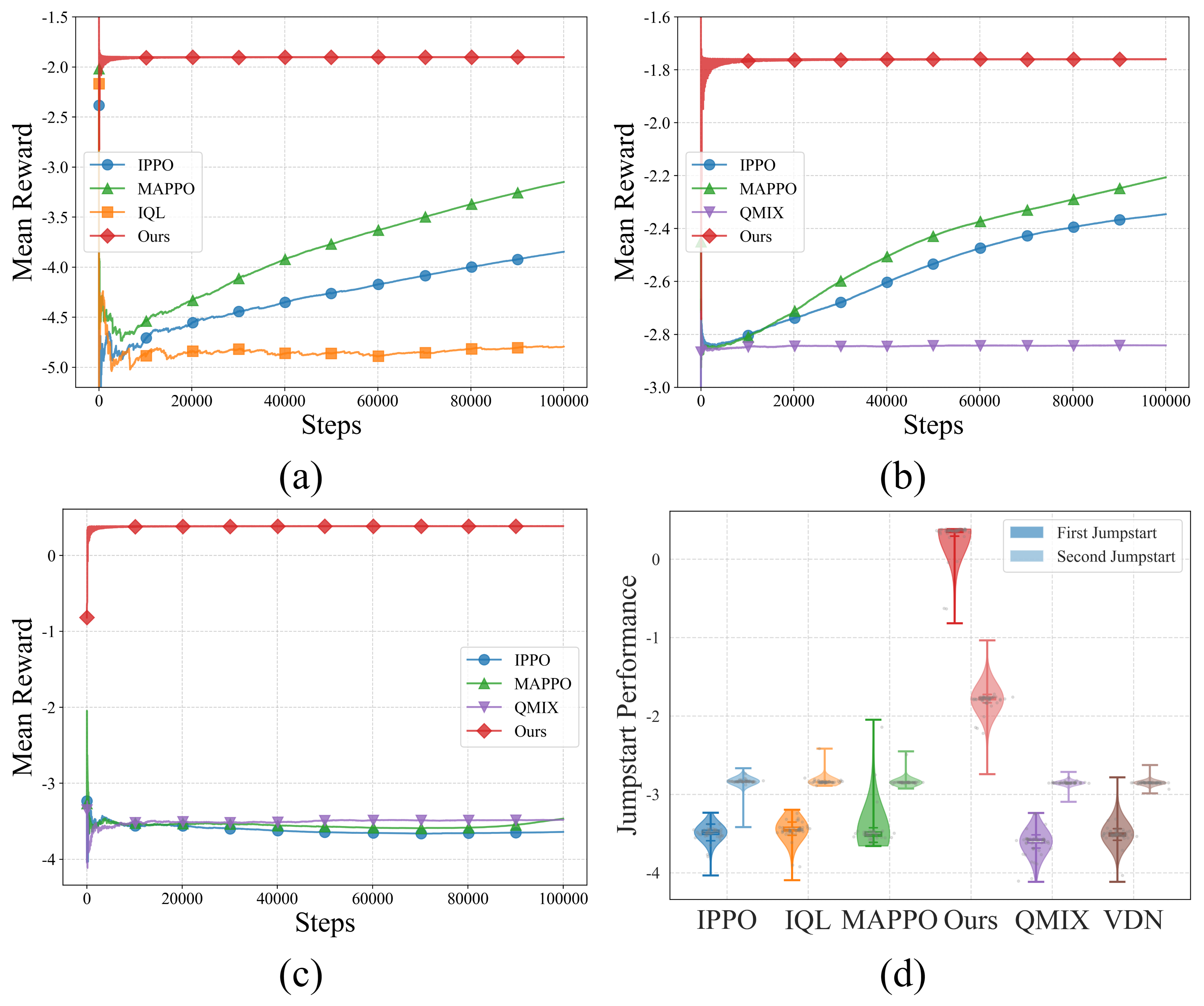} 
	\caption{Figure (a-c) present the experimental results when the test scenario is dynamically switched from Sce4 (a) to Sce5 (b) and Sce6 (c), showing the variation of average cumulative rewards over decision steps. Figure (d) illustrates the jumpstart performance of each algorithm during scenario switching, highlighting their ability to adapt quickly to new environments.}
	\label{fig:policy_line}
\end{figure}

As shown in Fig.\ref{fig:policy_line}(a)-(c), during the dynamic transition of the attack strategy, the average cumulative reward of our method consistently and significantly outperforms the baseline algorithms. \mrevise{The jumpstart results in Fig.\ref{fig:policy_line}(d) shows that our method achieves the highest average jumpstart value during both dynamic transitions.} \mrevise{This validates that the framework, through its heterogeneously separated pre-trained lower-layer RL agents (see Section \ref{sec:hsp}), can dynamically combine diverse defense strategies via the LLM's tactical planning, thereby handling unknown and evolving attack policies (A3).}

\subsection{RQ4: Robustness against Evolving Attack Scale}

To validate the proposed method's adaptability to dynamic attack scale, this section configures experimental scenarios that transition from Sce1 to Sce2 and Sce3. This setup increases the number of attackers from 6 to 8, thereby characterizing the growth in attack scale by escalating concurrent threats.

\begin{figure}[htbp]
	\centering
	\includegraphics[width=\columnwidth]{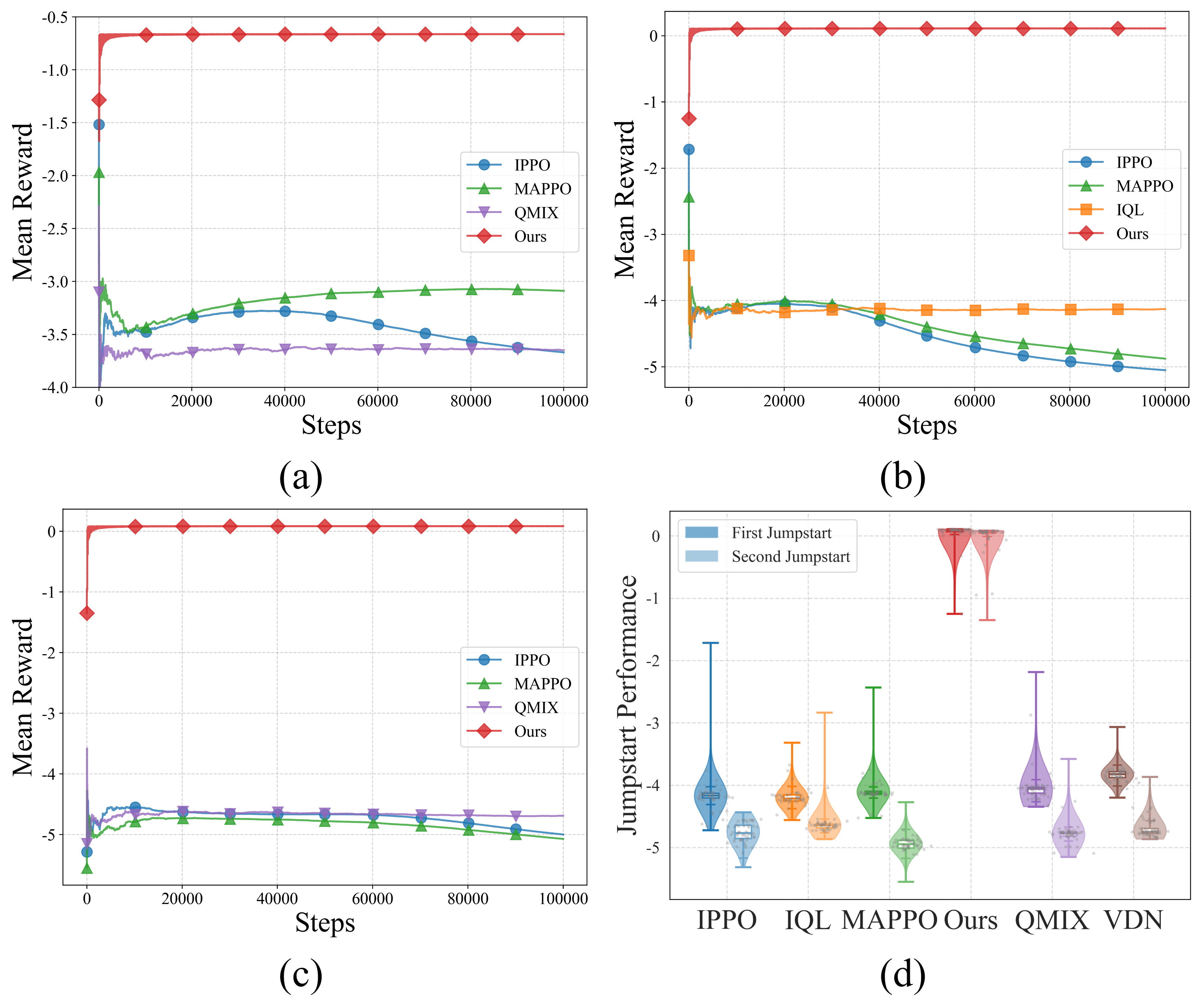} 
	\caption{Figure (a-c) present the experimental results when the test scenario is dynamically switched from Sce1 (a) to Sce2 (b) and Sce3 (c), showing the variation of average cumulative rewards over decision steps. Figure (d) illustrates the jumpstart performance of each algorithm during scenario switching, highlighting their ability to adapt quickly to new environments.}
	\label{fig:att_line}
\end{figure}

\mrevise{As can be seen from Fig.\ref{fig:att_line}(a)-(c), the proposed method consistently maintains a leading position in terms of average cumulative reward.} Particularly in Sce3, where the number of attackers doubles that of the defenders, baseline algorithms such as IPPO and IQL exhibit a notable decline in cumulative reward, while the proposed method still sustains a relatively high reward level. The jumpstart results in Fig.\ref{fig:att_line}(d) indicate that the proposed method achieves the highest jumpstart values during both transitions in attack scale, \mrevise{demonstrating its capability to quickly adapt to dynamic attack scale (A4).}

\revise{To investigate whether the variations in attack policies affect the robustness of defense frameworks under dynamically changing attack scales, we extend the aforementioned experiments that dynamically shifted attack scales across scenarios Sce1, Sce2, and Sce3. Specifically, we further replace the base attack policy (Recon) with the more threatening Penetrate and Impact policy, respectively, to examine the trends in jumpstart performance for different algorithms as the attack scale increases under these advanced adversarial behaviors. Experimental results are presented in Table~\ref{tab:jumpstart}. }

\begin{table}[htbp]
	\centering
	\caption{\revise{Jumpstart under Various Attack Policies and Scales.}}
	\label{tab:jumpstart}
	
	\newcolumntype{Y}{>{\centering\arraybackslash}X}
	
	\begin{tabularx}{\columnwidth}{lYYYY}
		\toprule
		\multirow{6}{*}{Algorithm} & \multicolumn{4}{c}{\textbf{Attack Strategy}} \\
		\cmidrule(lr){2-5}
		& \multicolumn{2}{c}{\textit{Penetrate}} & \multicolumn{2}{c}{\textit{Impact}} \\
		\cmidrule(lr){2-3} \cmidrule(lr){4-5}
		& \multicolumn{4}{c}{\textbf{Attack Scale}} \\
		\cmidrule(lr){2-5}
		& $6 \to 7$ & $7 \to 8$ & $6 \to 7$ & $7 \to 8$ \\
		\midrule
		IPPO  & -4.42 & -5.56 & -5.95 & -6.51 \\
		IQL   & -4.38 & -4.91 & -5.48 & -5.87 \\
		MAPPO & -4.24 & -4.95 & -5.12 & -5.89 \\
		QMIX  & -4.55 & -5.62 & -6.13 & -6.73 \\
		VDN   & -4.48 & -5.47 & -5.85 & -6.92 \\
		\midrule
		\textbf{Ours} & \textbf{0.12} & \textbf{-0.24} & \textbf{-2.05} & \textbf{-2.22} \\
		\bottomrule
	\end{tabularx}
\end{table}

\revise{The data analysis reveals that regardless of how the attack policy evolves, a dynamic increase in attack scale leads to a decline in the jumpstart performance of all baseline algorithms. In contrast, the proposed CyberOps-Bots framework consistently maintains the best jumpstart performance under all tested conditions. This finding indicates that the evolution of attack policy further impairs the ability of baseline algorithms to adapt to dynamic attack scales. The proposed framework, however, suffers relatively less from this effect due to its long-term memory mechanism, which enables continuous tracking and analysis of multiple attack chains. Consequently, even when confronted with complex threats characterized by the dual dynamics of evolving strategies and scales, the framework retains high robustness and rapid adaptation capability.}

\subsection{RQ5: Cyber Resilience through Balancing Defense Effectiveness and Availability}

This set of experiments validates the proposed framework's capability to balance long-term defense effectiveness with network availability protection during dynamic network changes. We analyze the relationship between the Maximum Episode Length and the Mean Healthy Ratio for all algorithms across the dynamic scenarios described in the preceding RQs.

\begin{figure*}[htbp]
	\centering
	\includegraphics[width=0.8\textwidth]{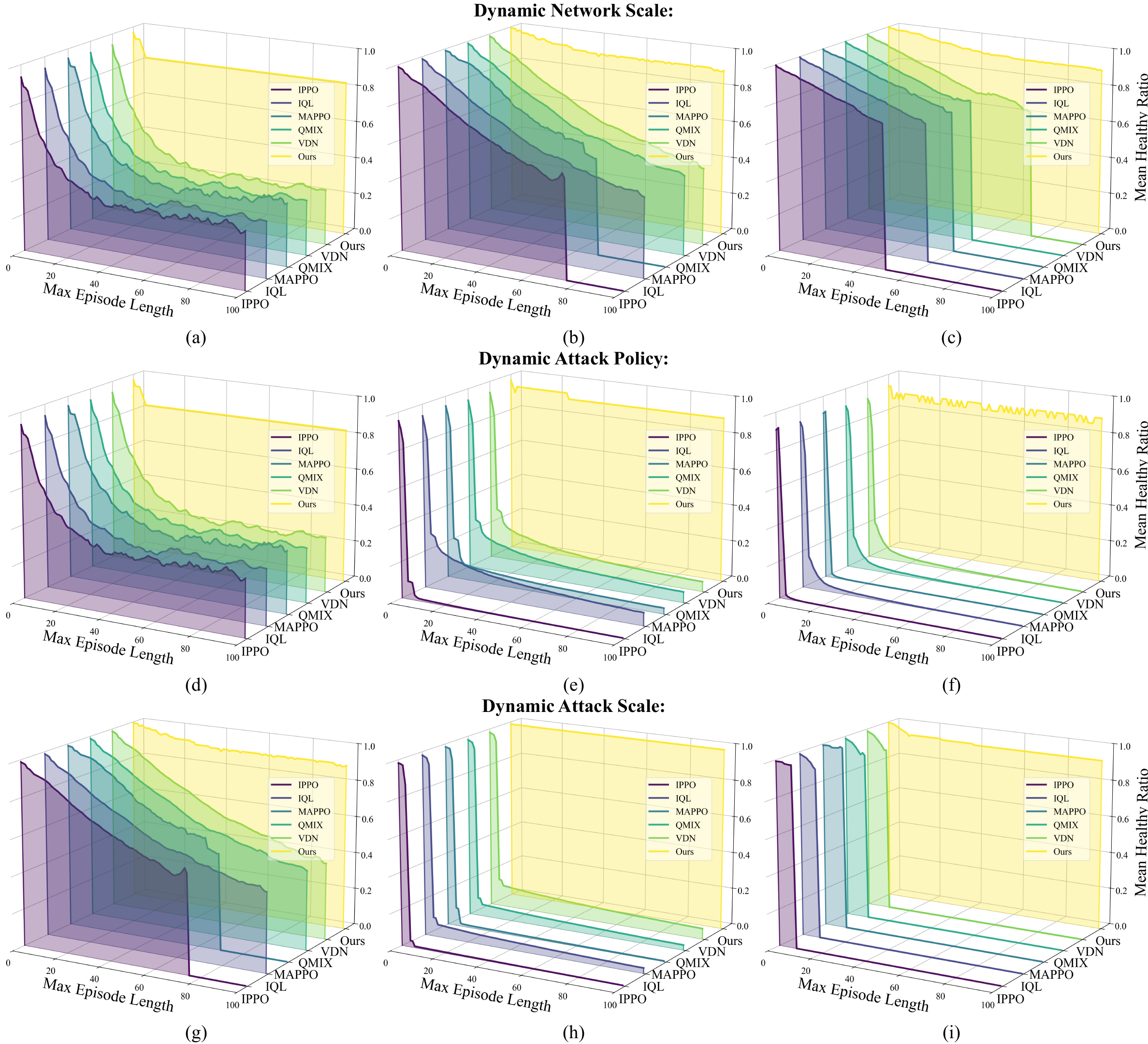}
	\caption{Comparison of the trade-off between defense persistence (Maximum Episode Length) and resilience (Mean Healthy Ratio) across different algorithms under three aspects: (a)-(c) dynamic network scale changes; (d)-(f) varying attack policies; and (g)-(i) increasing attack scales.}
	\label{fig:waterfall}
\end{figure*}

Specifically, Fig.\ref{fig:waterfall}(a-c) presents the results when the network scale dynamically expands from Sce1 (30 nodes) to Sce4 (150 nodes) and Sce7 (450 nodes). Fig.\ref{fig:waterfall}(d-f) corresponds to scenarios where the attack strategy switches from Sce4 (recon) to Sce5 (penetrate) and Sce6 (impact). Fig.\ref{fig:waterfall}(g-i) reflects the algorithmic performance as the attack scale gradually increases from Sce1 (6 attackers) to Sce2 (7 attackers) and Sce3 (8 attackers).

As shown in Fig.\ref{fig:waterfall}(a-c), with dynamic changes in network scale, the maximum episode length of baseline algorithms decreases significantly. \mrevise{Furthermore, maintaining a longer episode often requires sacrificing network availability, as seen in Fig.\ref{fig:waterfall}(c). In contrast, our method maintains the highest episode length (100) and health ratio (average 0.87) throughout the dynamic transition from Sce1 to Sce7.}

\mrevise{As shown in Fig.\ref{fig:waterfall}(d-f), our method consistently achieves the highest episode length and a high network healthy ratio.} In comparison, baseline algorithms tend to over-isolate nodes when countering the penetrate strategy, leading to a substantial drop in \mrevise{healthy} ratio (Fig.\ref{fig:waterfall}(e)). Under the impact strategy (Fig.\ref{fig:waterfall}(f)), the episode length shortens significantly due to the inability to effectively block concentrated attacks on high-value nodes.

\revise{As depicted in Fig.\ref{fig:waterfall}(g-i), when the attack scale grows dynamically, our method maintains the highest episode length (100) and a relatively high average health ratio (average 0.84 in Sce3). Conversely, the performance of baseline algorithms declines markedly after the attack scale increases, with both their maximum episode length and health ratio severely degraded.}

\subsection{RQ6: Resilience Enhancement through Proactive Network Hardening}

Proactive hardening refers to the capability of defenders to not only execute effective defense strategies upon the occurrence of a network attack but also proactively patch the network's vulnerabilities based on the incident, thereby preventing future malicious activities. To verify whether each algorithm can effectively perform hardening measures while defending, we statistically analyzes the average network vulnerability (see Definition \ref{def:state}) at the end of each episode across all the aforementioned scenarios.

\begin{figure}[htbp]
	\centering
	\includegraphics[width=\columnwidth]{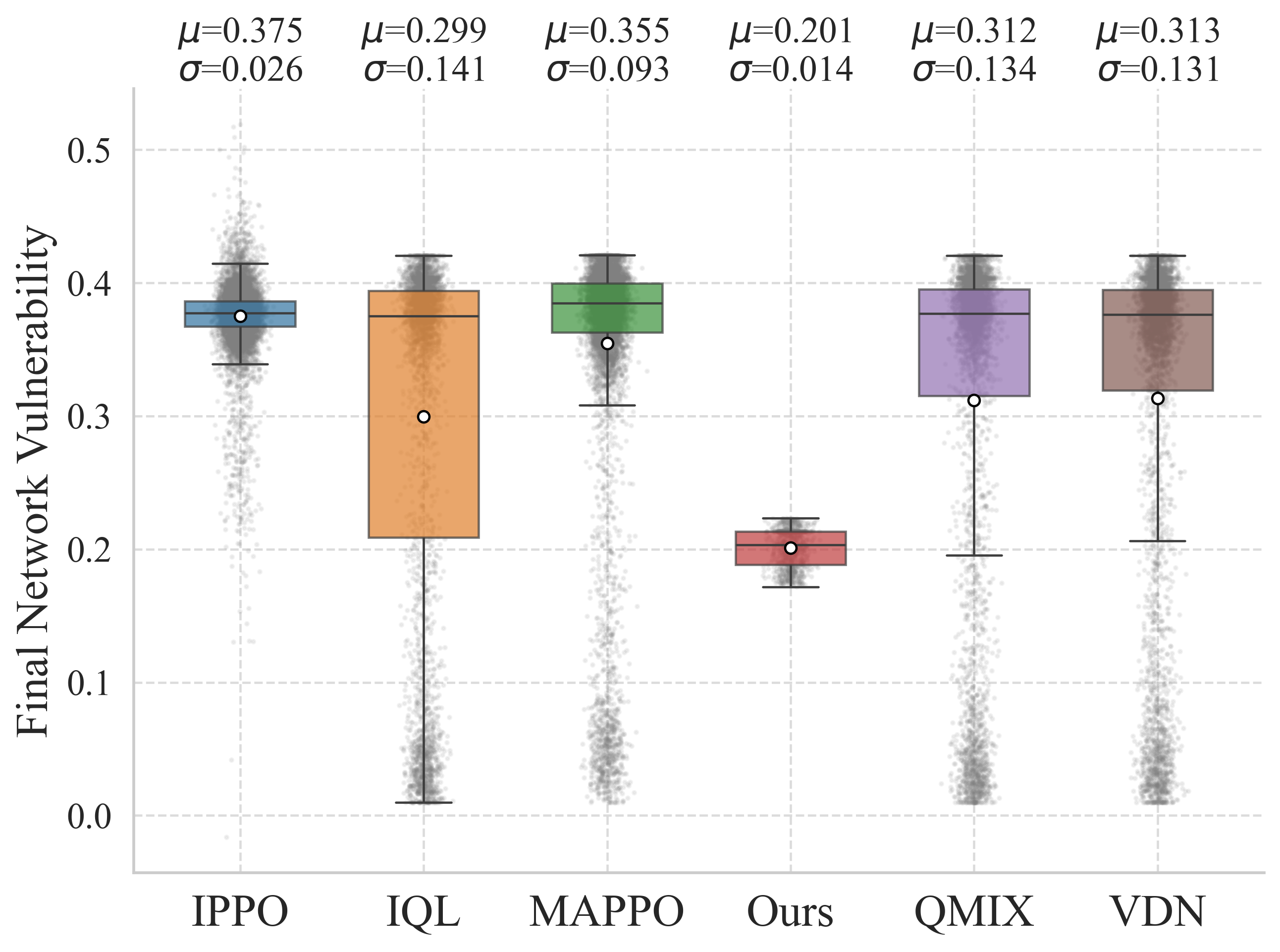} 
	\caption{This figure shows the average network vulnerability value at the end of each episode across all scenarios, reflecting the algorithms’ ability to defend against attacks while strengthening the network.}
	\label{fig:box}
\end{figure}

Fig.\ref{fig:box} illustrates the distribution of the average network vulnerability at the end of episodes for algorithms across all the experimental scenarios. The box corresponding to our method is positioned significantly lower than those of the baseline algorithms, and its distribution range is more concentrated. This indicates that our approach can consistently and stably reduce network vulnerability under dynamic conditions. In contrast, the vulnerability distributions of the baseline algorithms are more dispersed and contain more outliers, reflecting the instability of their strategies in achieving network hardening effects within dynamic environments. This result further validates that the CyberOps-Bots framework not only maintains immediate defense effectiveness against dynamic attack threats but also achieves long-term resilience reinforcement.

\subsection{RQ7: Operational Robustness Analysis regarding Efficiency and Reliability}

This section evaluates the computational overhead and decision reliability of the proposed framework. Table \ref{tab:efficiency} details the computational efficiency metrics across different network scales, including average token consumption and a contrast of our framework's decision time against the average time of all baseline algorithms. \mrevise{Table \ref{tab:reliability} compares the hallucination rates of the complete framework against ablation configurations. }

\begin{table}[tbp]
	\centering 
	\caption{Computational Efficiency of the LLM Agent.}
	\label{tab:efficiency}
	\begin{tabular}{ccccc}
		\toprule 
		\multirow{2}{*}{Scenario} & \multirow{2}{*}{Node Scale} & \multirow{2}{*}{Token(K)/Step} & \multicolumn{2}{c}{Time(ms)/Step} \\
		\cmidrule(lr){4-5} 
		&                             &                                & Ours & Baselines (AVG) \\
		\midrule 
		Sce1 & 30  & 1.64 & 335.12 & 72.09 \\
		Sce4 & 150 & 2.51 & 347.85 & 97.91 \\
		Sce7 & 450 & 3.62 & 366.43 & 141.54 \\
		\bottomrule 
	\end{tabular}
\end{table}

\begin{table*}[tbp]
	\centering
	\caption{\mrevise{Reliability Analysis of the Framework vs. Ablated Variants.}}
	\label{tab:reliability}
	\begin{tabular}{l c c c c}
		\toprule
		\multirow{2}{*}{Framework Name} & \multicolumn{2}{c}{Hallucination Rate} & \multicolumn{2}{c}{Defense Effectiveness} \\
		\cmidrule(lr){2-3} \cmidrule(lr){4-5}
		& Reasoning & Final Answer & Mean Episode Length & Mean Healthy Ratio \\
		\midrule
		Ours           & 8.90\%  & \textbf{0.71\%} & \textbf{98.91} & \textbf{89.72\%} \\
		Ours w/o ReAct & 8.52\%  & 8.52\% & 87.13 & 83.95\% \\
		Ours w/o STM   & 10.03\% & 1.64\% & 84.22 & 81.04\% \\
		Ours w/o LTM   & 8.91\%  & \textbf{0.71\%} & 66.87 & 69.08\% \\
		\bottomrule
	\end{tabular}
\end{table*}

\revise{In this experiment, the LLM’s ``hallucination’’ is defined based on the potential risks it introduces during the decision-making process, including two categories: The first category is factual hallucination, where the content generated by the LLM contradicts the specific cloud network context provided by the perception module. The second category is formatting hallucination, where the LLM’s output fails to strictly adhere to formatting requirements or tool-calling interface specifications. This includes errors in parameter types, out-of-range values, or structures incompatible with executable command requirements. Our formatting requirements for output align with standards of the OpenAI API formatting guidelines\cite{openai_introducing_2024}. The hallucination rate, which quantifies decision-making reliability, is calculated by counting the occurrences of the aforementioned issues over a defined number of decision steps.}

\mrevise{All data are averaged over 1000 decision steps across all scenarios in Table \ref{tab:senarios}.}

Experimental results indicate that token consumption and decision time are primarily correlated with network scale. As shown in Table~\ref{tab:efficiency}, token usage increases linearly as the network scale doubles. Given the context window capabilities of current mainstream LLMs, the framework demonstrates robust token efficiency and scalability. Regarding decision latency, by leveraging EAGLE-3 \cite{EAGLE} for inference acceleration and imposing strict constraints on LLM reasoning and output formats, our framework achieves millisecond-level decision times per step. Although this inference speed is naturally slower than lightweight RL baseline algorithms, the latency remains within an acceptable range considering the framework's superior performance. Notably, as the network scale expands, the time consumption of baseline algorithms exhibits a sharper increase due to the escalating computational complexity caused by the multiplication of input states. In contrast, our framework's decision time increases only marginally. This is because the increased scale primarily translates to more input tokens—which remain far below the maximum context limit—while the strict prompt constraints ensure that the time required for generation remains largely independent of network scale.

\begin{figure*}[htbp]
	\centering
	\includegraphics[width=0.8\textwidth]{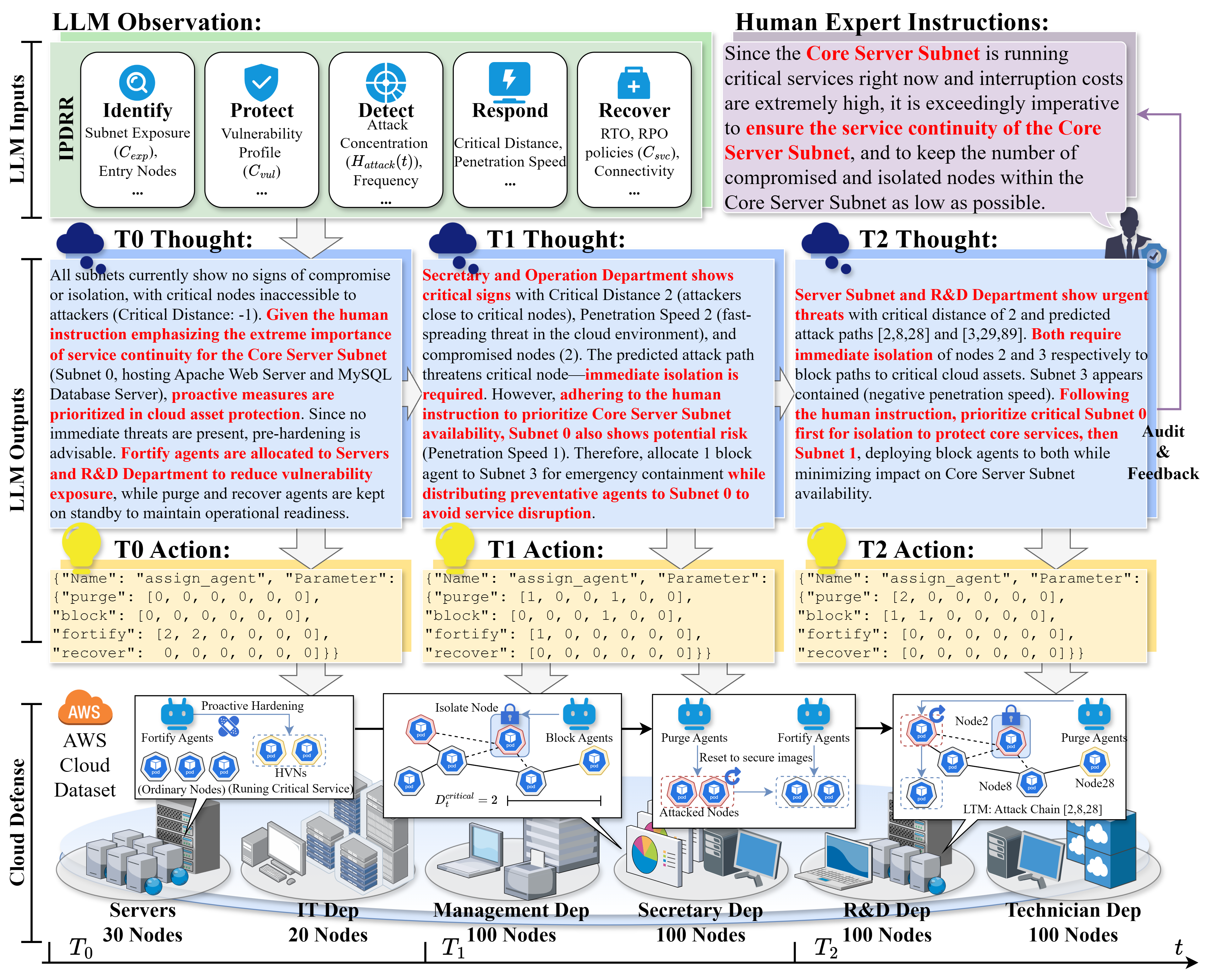}
	\caption{Illustration of the LLM agent's tactical evolution under human intervention. The diagram displays the reasoning chain and defense actions across three time steps (T0, T1, and T2) in response to expert instructions.}
	\label{fig:case}
\end{figure*}

\mrevise{As presented in Table~\ref{tab:reliability}, while the framework exhibits an 8.90\% hallucination rate during the reasoning process, the multi-step verification mechanism of the ReAct paradigm significantly reduces the final decision hallucination rate to 0.71\%.  When the ReAct module is removed, although the single-step reasoning hallucination drops slightly (8.52\%), the lack of iterative correction causes the final answer hallucination rate to spike, exceeding that of the complete framework. Furthermore, removing the STM module leads to a loss of decision context, increasing the reasoning hallucination rate to 10.03\%. While removing the LTM module has a negligible impact on hallucination rates, it substantially undermines defense effectiveness, as evidenced by the significant drops in both episode length and healthy ratio. In conclusion, the ReAct planning and memory mechanisms significantly improve the decision reliability of the framework.}

\subsection{RQ8: Support Capability for Human-in-the-Loop}

\revise{To evaluate the functional efficacy of the HITL mechanism, this section injects security expert instructions to test the framework's adaptability to new tactical objectives, analyzing how human intervention shifts the agent's decision-making logic. We also analyze the interpretability and auditability of the LLM agent's reasoning chain to evaluate the transparency of human-machine collaborative decision-making.}

By default, the defense strategy prioritizes high-value asset protection, treating the service continuity of all subnets with equal importance. \revise{In this experiment, the expert instruction in Fig.\ref{fig:case} was injected into the LLM agent via the perception module to override the default priorities.}

This instruction imposes a higher-level objective over the predefined reward function, requiring the framework to prioritize the availability of the server subnet. Upon receiving this instruction, the LLM generated reasoning chain shown in Fig.\ref{fig:case} over three time steps.

The sequential reasoning chain across these three time steps demonstrates the tactical evolution of the LLM agent under human intervention:
\begin{itemize}
	\item \textbf{T0 (Preventive Defense):} Prioritizes hardening the core server subnet in the absence of immediate threats.
	\item \textbf{T1 (Threat Response):} Balances multiple threats while maintaining core service continuity as the primary objective.
	\item \textbf{T2 (Precise Isolation):} Executes targeted isolation based on predicted attack paths, minimizing service interruption costs while conducting image reset and recovery.
\end{itemize}

Experimental results indicate that the LLM agent effectively comprehends and executes high-level objectives prioritizing core server continuity. 

\revise{Following the human instruction, the reasoning chain at T1 and T2 shows a deliberate trade-off: the agent accepted a higher risk in other subnets to ensure absolute availability of the core server subnet. This alignment between human intent and the agent’s subsequent reasoning logic provides empirical evidence of the HITL mechanism’s functional effectiveness.}

\begin{table*}[t]
	\centering
	\caption{\mrevise{Quantitative Evaluation of HITL Effectiveness on Defense Decision-Making.}}
	\label{tab:human_instructions}
	\begin{tabular}{l c c c}
		\toprule
		\multirow{2}{*}{Configuration} & Core Server Subnet & \multicolumn{2}{c}{Overall Metrics} \\
		\cmidrule(lr){2-2} \cmidrule(lr){3-4}
		& Subnet Mean Healthy Ratio & Mean Healthy Ratio & Mean Episode Length \\
		\midrule
		With Human Instructions    & \textbf{91.07\%} & \textbf{90.36\%} & 95.88 \\
		Without Human Instructions & 84.83\% & 89.72\% & \textbf{98.91} \\
		\bottomrule
	\end{tabular}
\end{table*}

\mrevise{To evaluate the practical efficacy of the HITL mechanism in guiding defense decisions, we report average results from 1,000 decision steps across all scenarios listed in Table \ref{tab:senarios}. The evaluation metrics encompass the healthy node ratio of the core server subnet specified by the human instruction, as well as the overall network-wide healthy node ratio and average episode length. The quantitative results are summarized in Table \ref{tab:human_instructions}. The analysis reveals that after injecting the human instruction to ``prioritize the availability of the server subnet'', the framework achieves a significantly higher healthy node ratio for the targeted subnet, demonstrating a clear improvement over the baseline without instruction. This indicates that the upper-layer LLM agent can accurately comprehend the high-level tactical intent conveyed via natural language and dynamically adjust its defensive focus accordingly. It is noteworthy that while executing this focused adjustment, the framework's overall defensive efficacy remains intact. The results confirm that the designed HITL mechanism can effectively and flexibly integrate human expert knowledge and instructions into the automated decision cycle.}

\revise{Furthermore, the ReAct-based reasoning log explicitly records the rationale behind tactical decisions and the integration of human instructions. This transparency provides security experts with an audit trail for real-time supervision and post-incident review, thereby enhancing the reliability and trustworthiness of human-machine collaboration in complex cloud defense scenarios. Crucially, these HITL auditing and inspection procedures provide essential oversight, effectively mitigating potential risks caused by LLM hallucinations.}

\section{Conclusion}

This paper proposes CyberOps-Bots, a hierarchical multi-agent RL framework for cloud network resilience. By synergizing LLM-based high-level planning with atomic RL execution, the framework effectively addresses critical adaptability challenges in dynamic network environments. Experimental results demonstrate a 68.5\% improvement in network availability and a 34.7\% jumpstart gain without retraining. Notably, this is the first framework to offer inherent Human-in-the-Loop support and superior interpretability. \revise{While the centralized LLM architecture poses potential scalability and security constraints, future work will focus on decentralized multi-LLM coordination, integrating model security safeguards, and designing cross-layer decision verification mechanisms to enhance the security of the autonomous defense ecosystem. Furthermore, investigating automated prompt optimization and sensitivity analysis will be prioritized to further bolster the framework's robustness against semantic variations.} Overall, CyberOps-Bots represents a significant advancement toward autonomous, adaptive, and sustainable cloud defense.

\section*{Acknowledgments}
This work was supported in part by the National Natural Science Foundation of China under Grants 61902427 and 62471064, and by the National Key Research and Development Program of China under Grants 2023YFC3306305 and 2022YFB2902200.

\bibliographystyle{IEEEtran}
	
\bibliography{ref.bib}

\vspace{-3em}
\begin{IEEEbiographynophoto}{Yixiao Peng}
	received the B.E. degree in information countermeasure technology from Information Engineering University, Zhengzhou, China, in 2024, where he is currently pursuing the M.D. degree in cyberspace security. His research interests include network security and large language models.
\end{IEEEbiographynophoto}
\vspace{-3em}
\begin{IEEEbiographynophoto}{Hao Hu}
	received his Ph.D. degree in the Zhengzhou Information Science and Technology Institute in 2018. He has been an associate professor with the National Digital Switching System Engineering and Technological R\&D Center since 2018. His research interests include attack-defense modeling and proactive defense.
\end{IEEEbiographynophoto}
\vspace{-3em}
\begin{IEEEbiographynophoto}{Feiyang Li}
	received the B.E. degree in information countermeasure technology from Information Engineering University, Zhengzhou, China, in 2023. He is currently pursuing the Ph.D. degree in cyberspace security. His research interests include reinforcement learning and automated penetration testing.
\end{IEEEbiographynophoto}
\vspace{-3em}
\begin{IEEEbiographynophoto}{Xinye Cao}
	received the B.E. degree in electronic and information engineering from Central China Normal University (CCNU), Wuhan, China, in 2020. She is currently pursuing the Ph.D. degree in cyberspace security with the National Engineering Research Center for Mobile Network Technologies, Beijing University of Posts and Telecommunications (BUPT). Her research interests include large AI models for future wireless communication systems and wireless communication security.
\end{IEEEbiographynophoto}
\vspace{-3em}
\begin{IEEEbiographynophoto}{Yingchang Jiang}
	is a Ph.D. candidate at the Information Engineering University, Zhengzhou, China. His main research interest is to analyze cyber threat intelligence using LLMs and deep learning techniques.
\end{IEEEbiographynophoto}
\vspace{-3em}
\begin{IEEEbiographynophoto}{Jipeng Tang}
	received the B.M. degree in Management Science and Engineering from Information Engineering University, Zhengzhou, China, in 2024, where he is currently pursuing the M.D. degree in cyberspace security. His research interests include network security and knowledge graphs.
\end{IEEEbiographynophoto}
\vspace{-3em}
\begin{IEEEbiographynophoto}{Guoshun Nan}
	is currently a Full Professor at the National Engineering Research Center for Mobile Network Technologies, Beijing University of Posts and Telecommunications, Beijing, China. He has a broad interest in multimodal learning, large language models, and 6G network security.
\end{IEEEbiographynophoto}
\vspace{-3em}
\begin{IEEEbiographynophoto}{Yuling Liu}
	received his Ph.D. degree from the Institute of Software, Chinese Academy of Sciences in China. He is currently a senior engineer at the Institute of Information Engineering, Chinese Academy of Sciences. His research areas are network security situational awareness, network security, big data analysis, and security measurement and certification.
\end{IEEEbiographynophoto}

\end{document}